\documentclass[aps,prd,twocolumn,preprintnumbers,superscriptaddress,10pt,nofootinbib]{revtex4-1}
\usepackage{epsf,color,amsmath,amssymb,amsfonts}
\usepackage{enumerate}
\usepackage{hyperref}
\usepackage{graphicx}
\usepackage{psfrag}
\usepackage{dcolumn,bm}

\newcommand{\nc}{\newcommand}
\nc{\rnc}{\renewcommand }
\nc{\tep}{\tilde{\epsilon}}
\rnc{\d}{\mathrm{d}}
\nc{\D}{\partial}
\nc{\bg}{\bar{g}}
\nc{\g}{\gamma}
\rnc{\o}{\omega}
\nc{\n}{{(n)}}
\rnc{\t}{\tau}
\nc{\ep}{\epsilon}

\def\ba{\begin{eqnarray}}
\def\ea{\end{eqnarray}}
\def\be{\begin{equation}}
\def\ee{\end{equation}}
\def\nn{\nonumber}

\newcommand{\DD}{\mathcal{D}}
\newcommand{\LL}{\mathcal{L}}

\newcommand{\HH}{\mathcal{H}}
\newcommand{\RR}{\mathbb{R}}
\newcommand{\TT}{\mathcal{T}}
\newcommand{\KK}{\mathcal{K}}
\newcommand{\OO}{\mathcal{O}}
\newcommand{\nud}{_{\nu}}

\newcommand{\muu}{^{\mu}}

\newcommand{\PP}{\mathcal{P}}

\begin{document}
\onecolumngrid

\title{Towards a general fluid/gravity correspondence}

\author{Natalia Pinzani-Fokeeva}
\email{N.PinzaniFokeeva@uva.nl}
\affiliation{Instituut voor Theoretische Fysica, Universiteit van Amsterdam \\
Science Park 904, 1090 GL Amsterdam, The Netherlands}

\author{Marika Taylor}
\email{M.M.Taylor@soton.ac.uk}
\affiliation{Mathematical Sciences and STAG research centre, University of Southampton, Highfield, Southampton SO17 1BJ, U.K.} 

\begin{abstract}
We set up the construction of generic $(d+2)$-dimensional metrics corresponding to $(d+1)$-dimensional fluids, representing 
holographically the hydrodynamic regimes of the putative dual theories. We give general seed equilibrium metrics appropriate to generic bulk
stress energy tensors and discuss the implications of conformal rescalings of the hypersurface on which the fluid is defined. We then show how to obtain the corresponding hydrodynamic metrics using a relativistic gradient expansion and discuss the integrability conditions of the resulting equations. The stress energy tensors of the resulting fluids, both at and away from equilibrium,  satisfy a quadratic constraint. We interpret this constraint in terms of two possible equations of state for the fluid and show that only one of the two equations is physical. We illustrate our discussions with the example of the cutoff AdS fluid, for which 
we find the precise interpretation in terms of deformations of the UV conformal field theory. Finally we discuss the relation between the modern fluid/gravity approach taken in this paper and the earlier membrane paradigm. 
\end{abstract}

\maketitle

\section{Introduction}
The holographic principle proposes an equivalence between $(d+2)$-dimensional  gravitational theories and $(d+1)$-dimensional quantum field theories.
A concrete realization of holography in which the dictionary between bulk and boundary data has been extensively developed is the case of  asymptotically locally Anti-de Sitter spacetimes (AdS) corresponding to conformal field theories (CFT). 
However, if holography is the correct paradigm,  holographic descriptions should exist  for more general gravitational theories with different asymptotics. Holography has in recent times been pushed beyond AdS asymptotics, particularly to non-relativistic dualities such as
Lifshitz and Schrodinger, but constructing a holographic duality for even asymptotically flat spacetimes has been a long standing challenge. 

A generic feature of any quantum field theory is the existence of a hydrodynamic description capturing the long-wavelength behaviour of the microscopic degrees of freedom near to thermal equilibrium. If holography is correct, such regimes
have to be reproduced within the context of the holographic duality.
The fluid/gravity correspondence \cite{Bhattacharyya:2008jc} exactly realises this scenario: nearby  gravitational solutions describing the hydrodynamic regime have been explicitly constructed for AdS black brane solutions, which correspond to a conformal fluid at thermal equilibrium. 
This framework was later extended to classes of non-conformal fluids, see for example
\cite{Kanitscheider:2009as,David:2009np}; \cite{Rangamani:2009xk} reviews the many other extensions of the fluid/gravity relation. 

An extension of fluid/gravity duality to vacuum Einstein gravity has also been developed over the past few years.
In \cite{Bredberg:2010ky,Bredberg:2011jq, Compere:2011dx}, the incompressible Navier-Stokes equations in $(d+1)$ dimensions were shown to be dual to a $(d+2)$-dimensional Ricci-flat metric. This holographic duality for Ricci flat spacetime has attracted much interest and the correspondence has been extended in several directions. 
A systematic construction of the metric to all orders in the hydrodynamic expansion was provided in \cite{Compere:2011dx}; this determines the corresponding specific corrections to the incompressible Navier-Stokes equations.  Subsequently in \cite{Compere:2012mt,Eling:2012ni} the fluid/gravity relation was reformulated in a manifestly relativistic expansion.  
The thermal state corresponds to the Rindler spacetime and near equilibrium solutions represent the hydrodynamic behaviour of a dual fluid living on a finite cutoff time-like hypersurface $\Sigma_c$. In \cite{Meyer:2013sva}  relations among the second order transport coefficients have been derived and the authors of \cite{Chirco:2011ex, Eling:2012xa,Zou:2013ix} studied higher derivative corrections.
Extensions to spherical horizon topologies and to de Sitter have been studied in \cite{Bredberg:2011xw, Huang:2011he, Nakayama:2011bu, Anninos:2011zn} while charged fluids were explored in \cite{Niu:2011gu,Zhang:2012uy,Lysov:2013jsa}. The role of Petrov conditions has been explored in \cite{Lysov:2011xx,Huang:2011he,Huang:2011kj,Cai:2013uye,Ling:2013kua} and solution generating symmetries were investigated in \cite{Berkeley:2012kz}. Other related works include \cite{Kuperstein:2011fn,Chapman:2012my,Matsuo:2012pi,Kuperstein:2013hqa,Mukhopadhyay:2013gja}.

In this paper we give a general prescription for constructing hydrodynamic solutions associated with a $(d+1)$-dimensional timelike flat hypersurface $\Sigma_c$ inside a $(d+2)$-dimensional bulk spacetime, which has a horizon and is supported by a generic bulk stress tensor. Note that this hypersurface does not necessarily need to be chosen to be near the horizon nor near the (conformal) boundary of the space-time.  
We begin in section \ref{two} by setting up an appropriate form for a bulk metric dual to an equilibrium fluid and its relation to the Bondi-Sachs form \cite{Bondi:1962px,Sachs:1962wk} as well as the metric ansatz used in holographic numerical simulations such as \cite{Chesler:2008hg}.

By construction, when the hypersurface $\Sigma_c$ is flat, the induced Brown-York stress energy tensor on this hypersurface takes a fluid form, and one can read off the defining properties of the putative dual fluid. We highlight the fact that, when the hypersurface is only conformally flat, as is indeed the case for AdS/CFT, the holographic stress tensor cannot  be the Brown-York stress energy tensor but rather should be 
conformal to the Brown-York stress energy tensor. We also emphasise the implications of the fact that fixing a Dirichlet Minkowski condition on $\Sigma_c$ generically leads to non-canonical normalisations of Killing vectors both at asymptotic infinity and at the horizon.

As noted in \cite{Compere:2012mt,Eling:2012ni}, the fluid stress energy tensor satisfies a quadratic constraint, which acts as an equation of state. This implies that, for any given data for the bulk stress energy tensor on $\Sigma_c$, there are two distinct bulk solutions. For vanishing bulk stress energy tensor these correspond to the Rindler fluid and the Taub fluid, respectively \cite{Compere:2012mt,Eling:2012ni}. The Taub fluid has strictly negative energy density and temperature and is hence unphysical. In section \ref{three} we consider various examples of bulk stress energy tensors, included cosmological constant and gauge fields and we discuss the interpretations of the positive and negative signs in the equation of state. Only one sign in the equation of state gives rise to a physical fluid; the opposite sign is always associated with a negative temperature and is therefore not physical. 

Using our examples, we observe that the existence of a flat timelike hypersurface is (as one would expect) highly non-trivial: in the absence of negative bulk curvature this requirement forces us into scaling regions of black holes with spherical horizon topologies. We also note that in the presence of generic matter the Einstein equations no longer form a nested hierarchy of decoupled equations, as they do in the vacuum and cosmological constant cases, see  \cite{Bondi:1962px,Sachs:1962wk} and \cite{Chesler:2008hg}. Since the equations do not decouple, even equilibrium (stationary) solutions are hard to find. 

In section \ref{four} we begin a preliminary investigation of the relation between the fluid stress energy tensor and the renormalised holographic stress energy tensor, the latter being defined in cases for which the holography duality is under control, e.g. asymptotically AdS or Lifshitz spacetimes. We use this perspective to understand why the fluid stress energy tensor for conformally flat hypersurfaces must be defined in terms of the conformally rescaled Brown-York stress energy tensor. 

In section \ref{five} we provide a general set up for hydrodynamics by promoting the defining parameters of the equilibrium bulk solutions to be position dependent. Working in a relativistic gradient expansion, we work out the hydrodynamic equations of motion. We show that the conservation of the dual fluid is associated with the integrability of the bulk equations and we derive a general expression for the first order dissipative corrections to the fluid stress energy tensor. 

As an example of our formalism we revisit the case of cutoff AdS. We compute the first order hydrodynamic metric corresponding to a Dirichlet condition on the hypersurface $\Sigma_c$. As this hypersurface is taken towards the conformal boundary we recover the results of \cite{Bhattacharyya:2008jc}. For a generic choice of $\Sigma_c$, we work out the asymptotic expansion of the metric in the neighbourhood of the conformal boundary. This metric remains asymptotically locally AdS but the background metric for the dual CFT is no longer conformally flat. We give the precise form for this background metric, up to first order in derivatives, thereby identifying the precise deformation of the dual CFT captured by the Dirichlet condition in the bulk. Earlier discussions of the mixed nature of the boundary condition at infinity can be found in 
\cite{Bredberg:2010ky,Heemskerk:2010hk,Faulkner:2010jy}, see also the related work \cite{Nickel:2010pr}. Discussions of the hydrodynamic behaviour on a finite cutoff hypersurface $\Sigma_c$ include \cite{Brattan:2011my,Emparan:2013ila};  see also \cite{Cai:2011xv,Bai:2012ci}.

The fluid/gravity duality is not the first time in which the physics of fluids has been linked to that of gravity. The membrane paradigm
was introduced by \cite{Damour:1979,Damour:1982,Price:1986} and postulates that any black hole horizon can be thought of as a membrane 
exhibiting fluid-like behaviour. In particular it was shown that certain components of Einstein equations could be recast in the form of dissipative non-relativistic Navier-Stokes equations. This fluid however has a negative bulk viscosity, a signal that this membrane fluid is unphysical, and
that the membrane paradigm itself should perhaps only be considered as a formal (if convenient) treatment of horizon dynamics.

A natural question is what is the relation between the membrane paradigm and modern fluid/gravity approaches.
It was already noted  in \cite{Hubeny:2009zz} that in fluid/gravity duality the dynamics of the entire spacetime is encoded holographically by the fluid associated with the boundary of the spacetime, while the membrane fluid is encoded on the horizon,
see also the related works \cite{Eling:2009sj,Eling:2009pb,Eling:2010yv,Kovtun:2003wp}. An extension of the membrane paradigm to hypersurfaces in the interior of the spacetime has been studied in \cite{Iqbal:2008by}, and recent works attempting to find connections between the membrane paradigm and fluid/gravity by studying hydrodynamic behaviour on a finite cutoff hypersurface $\Sigma_c$ include \cite{Brattan:2011my,Emparan:2013ila}, as well as \cite{Cai:2011xv,Bai:2012ci}.

In section \ref{six} we discuss the relation between the fluid/gravity approach and the membrane paradigm, by developing the work of
\cite{Gourgoulhon:2005ch}. The latter generalized the Damour equations to the case 
of general hypersurfaces foliating the bulk spacetime, with the hypersurfaces not necessarily being null and not  being assumed to be close to the horizon. In particular, \cite{Gourgoulhon:2005ch} constructed a generalised Damour-Navier-Stokes equation on codimension one timelike hypersurfaces. We extend the work of \cite{Gourgoulhon:2005ch} to other components of the Einstein equations and then use this setup
to compare the membrane approach with the fluid/gravity construction. We highlight the different identifications of both fluid parameters and the associate transport coefficients in the two approaches. In section \ref{seven} we conclude, while the appendices contain detailed derivations of certain results used in the main text. 

\section{Generic fluids in Einstein gravity } \label{two}

\subsection{The general seed metric ansatz}

We are interested in studying the hydrodynamics of fluids associated with a $(d+1)$-dimensional
timelike flat hypersurface $\Sigma_c$ foliating a $(d+2)$-dimensional bulk spacetime. 
We begin by  considering a generic static solution; having constructed static seed solutions one can always boost them to obtain stationary solutions corresponding to fluids with non-zero velocity. A convenient metric ansatz in Eddington-Finklestein type coordinates is of the form
\be \label{ansatz}
ds^2 = 2 dt dr - f(r) dt^2 + g(r) dx^{i} dx_{i}, 
\ee
with $i=1,\dots d$. On  hypersurfaces $\Sigma_c$ of constant $r=r_c$
the induced metric is therefore
\be \label{induced}
ds^2|_{\Sigma_c}= \gamma_{ab}dx^adx^b= - f(r_c)dt^2+g(r_c) dx_i dx^i,
\ee
where $x^a=(t,x^i)$. We assume that the hypersurface under consideration has a non-degenerate metric, i.e. $f(r_c)$, $g(r_c) \neq 0$, and that there is a horizon at $r = r_{H} \neq r_c$ where $f(r_H) = 0$. Each hypersurface is a worldline of observers with constant acceleration
\be
a=\frac{1}{2}\frac{f^\prime}{\sqrt{f}},
\ee
where the prime denotes a radial derivative. 
The temperature associated with the horizon is
\be
T=\frac{1}{4\pi}{f^\prime}(r_H) \label{temp}.
\ee

After  rescaling the coordinates  $t\rightarrow \bar{t}=\sqrt{f(r_c)}t$ and $x^i\rightarrow \bar{x}^i=\sqrt{g(r_c)}x^i$ the induced metric on $\Sigma_c$ can always be written in a manifestly flat form, i.e. the metric is
\be \label{static}
ds^2 = \frac{2 d \bar{t} dr}{\sqrt{f(r_c)}} - \frac{f(r)}{f(r_c)} d\bar{t}^2 + \frac{g(r)}{g(r_c)} d \bar{x}_i d \bar{x}^i. 
\ee
Boosting this metric results in 
\ba
ds^2 &=& - \frac{2}{\sqrt{f(r_c)}} u_a d \bar{x}^a dr + \frac{g(r)}{g(r_c)} d \bar{x}_a d \bar{x}^a + \nn\\
&&\quad+\left ( \frac{g(r)}{g(r_c)} -  \frac{f(r)}{f(r_c)} \right ) (u_a d \bar{x}^a)^2,
\ea
where the velocity is
\be
u^a = \gamma(1, v^i); \qquad \gamma = (1 - v^i v_i)^{-1/2},
\ee
as usual. The boost
manifestly preserves the induced metric on $\Sigma_c$. Defining
\be \label{redef1}
{G}(r) \equiv \frac{g(r)}{g(r_c)}; \qquad
{F}(r) \equiv \frac{f(r)}{f(r_c)}; \qquad \lambda \equiv \sqrt{f(r_c)}, 
\ee
such that ${G}(r_c) = {F}(r_c) = 1$, the metric can be written as
\be
ds^2 = - \frac{2}{\lambda} u_a d \bar{x}^a dr + {G}(r) d \bar{x}_a d \bar{x}^a + \left ( {G}(r) - {F}(r) \right ) (u_a d \bar{x}^a)^2. \label{form1}
\ee
The latter is a generic form for the equilibrium metric; in most of what follows we will work with the metric in this form and we will drop the barred notation for the rescaled coordinates. 
This is the most general seed equilibrium metric, with the corresponding fluid parameters (see below, \eqref{pressure} and \eqref{energy}) being $(u_a,G'(r_c),F'(r_c))$ where again the primes denote radial derivatives. Depending on the bulk stress energy tensor and matter present, the fluid may have other parameters. For example, if there is a bulk gauge field then the ansatz for the gauge field consistent with stationarity would be
\be
\underline{A} = (\mu(r) + (r-r_c) \rho(r)) u_a dx^a.
\ee
Then one would regard $\mu(r_c)$ as a boundary condition, characterising the chemical potential in the field theory, and $\rho(r_c)$ as characterising the charge density in the fluid. Not all of the fluid parameters are independent, since they are related by the Hamiltonian constraint, equivalent to the equation of state for the fluid, as well as conservation equations. For example, for a cosmological constant stress energy tensor, the equation of state implies that only one out of $F'(r_c)$ and $G'(r_c)$ is independent. 

\bigskip

The original metric ansatz \eqref{ansatz} is analogous to that used in \cite{Chesler:2008hg} and subsequent works, although in these papers $g(r)$ was rewritten as the square of a function, i.e. $g(r) = \Sigma(r)^2$. By rescaling the radial coordinate one can rewrite the metric in a static Bondi-Sachs \cite{Bondi:1962px,Sachs:1962wk} parametrisation:
\be
ds^2 = 2 e^{2 \beta(r_b)} dt dr_b - e^{2 \phi(r_b) + 4 \beta(r_b)} d t^2 + r_b^2 dx^i dx_i,
\ee
where $r_b$ is the new radial coordinate and $f(r) \equiv e^{2 \phi(r_b) + 4 \beta(r_b)}$ with $e^{2 \beta(r_b)} = 2 \sqrt{g} (\partial_r g)^{-1}$. This form of the metric is particularly convenient for solving the vacuum Einstein equations, as they form a nested hierarchy with the equation for $\beta$ solvable first, then $\phi$ may be found using the solution for $\beta$. In \cite{Chesler:2008hg}  the same nested hierarchy was found in the case of pure cosmological constant: again $g(r)$ could first be determined, with $f(r)$ then determined using the solution for $g(r)$. We will recover the same structure below but for a generic bulk stress energy tensor one cannot in general analytically integrate the equation for $g(r)$ even in static equilibrium situations.

\bigskip

Let us return for a moment to the original coordinates in \eqref{ansatz}. Clearly the induced metric on $\Sigma_c$ given in \eqref{induced} is conformal to a flat metric in which the effective speed of light is given by
\be
c_{l}^2 = \frac{f(r_c)}{g(r_c)}. 
\ee
Near the horizon, $f(r_c)$ is small and therefore the effective speed of light approaches zero. If one works with the original coordinates, rather than the rescaled coordinates, then the boost should preserve the induced metric on $\Sigma_c$. This implies that the boost must use $c_l$ as the effective speed of light, resulting in 
\ba
ds^2 &=& - 2 \frac{U_a}{c_l}  dx^a dr + g(r) (dx^i dx_i - c_l^2 dt^2) + \nonumber\\
&&\quad +(U_a dx^a)^2 \left (g(r) - \frac{f(r)}{c_l^2} \right ) \label{form2} 
\ea
with
\be
U_a = \gamma_l \left( c_l, \frac{v_i}{c_l}\right); \qquad 
\gamma_l^2 = (1 - c_l^{-2} v_i v^i)^{-1}. 
\ee
Defining $\hat{t} = c_l t$ this boosted metric becomes
\ba
ds^2 &=&  -2 \frac{\hat{U}_a}{c_l}  dx^a dr + g(r) (dx^i dx_i - d \hat{t}^2) +\nn\\
&&\quad+(\hat{U}_a dx^a)^2 \left (g(r) - \frac{f(r)}{c_l^2} \right ) \label{form3}
\ea
with 
\be
\hat{U}_a = \hat{\gamma}_l (1 , \hat{v}_i);  \qquad \hat{\gamma}_l^2 = (1 - \hat{v}^i \hat{v}_i)^{-1},
\ee
and $\hat{v}_i = dx^i/d \hat{t}$. 

In equilibrium the three forms of the metric \eqref{form1}, \eqref{form2} and \eqref{form3} differ from each other by trivial rescalings of the coordinates. Once one goes beyond equilibrium into the hydrodynamic regime, however, the three forms of the metric are no longer equivalent. The reason is that the metric functions $f(r)$ and $g(r)$ depend on the thermodynamic quantities: the temperature, charge etc. In the hydrodynamic regime the latter are promoted to be spatially dependent, and therefore both the conformal factor of $\Sigma_c$ in \eqref{form2} and the effective speed of the light $c_l$ on $\Sigma_c$ become spatially dependent. Using the coordinates of \eqref{form3}, the effective speed of light on the hypersurface does not vary along $\Sigma_c$ but the conformal factor of the induced metric is still (in general) becoming spatially dependent as one extends the solution into the hydrodynamic regime. By working with the coordinates \eqref{form1} one ensures that the induced metric on $\Sigma_c$ 
remains flat. In other words, this choice of coordinates is appropriate if one wants to impose a fixed Dirichlet boundary condition on $\Sigma_c$. 

\bigskip

It is useful to make one further rewriting of \eqref{form1} to obtain
\be
ds^2 = - \frac{2}{\lambda} u_a d {x}^a dr + {G}(r) h_{ab} d {x}^a d {x}^b - {F}(r) u_a u_b dx^a dx^b. \label{formf}
\ee
where $h_{ab} = \eta_{ab} + u_a u_b$, with 
the inverse metric being.
\be
g^{(0)rr}=\lambda^2 F;\qquad g^{(0)ra}=\lambda u^a;\qquad g^{(0)ab}=\frac{h^{ab}}{G}.
\ee
The Ricci tensor is then\footnote{Our convention for the Riemann tensor is $R^{\mu}_{\rho\nu\sigma}=\partial_{\nu}\Gamma ^{\mu}_{\rho\sigma}+\Gamma^{\mu}_{\lambda\nu}\Gamma^{\lambda}_{\rho\sigma}-(\nu \leftrightarrow\sigma)$. Details of the computation of the curvature at leading and subleading order may be found in the appendix.} 
\ba
R_{rr}^{(0)}&=&\frac{dG'^2}{4G^2}-\frac{dG''}{2G}, \\
\lambda^{-1}R_{ra}^{(0)}&=&u_a\left( \frac{1}{2}F''+\frac{dG'F'}{4G}\right),\nonumber\\
\lambda^{-2}R_{ab}^{(0)}&=&u_au_b\left( \frac{1}{2}FF''+\frac{dG'FF'}{4G} \right)+\nonumber\\
&&+h_{ab}\left( \frac{(2-d)}{4}\frac{G'^2F}{G}-\frac{1}{2}F'G'-\frac{1}{2}FG''    \right), \nonumber 
\ea
where in anticipation of extending to the hydrodynamic regime the superscript denotes that we are working to zeroth order in $x^a$ derivatives. 

We assume that the metric satisfies Einstein's equations with a given stress energy tensor, i.e. the Einstein tensor satisfies 
\be
G_{\mu \nu} = {\cal T}_{\mu \nu}.
\ee
Throughout this paper we will use ${\cal T}$ to denote the bulk stress energy tensor and reserve $T$ to denote the stress energy tensor associated with the fluid on $\Sigma_c$. 
Einstein's equations then read 
\ba\label{eqmotion0}
G''&=&\frac{1}{2}\frac{G'^2}{G}-\frac{2G}{d}\TT_{rr}^{(0)},\\
F'G'&=&-\frac{1}{2}(d-1)\frac{FG'^2}{G}+\frac{4GF}{d}\TT_{rr}^{(0)}+\frac{4G}{d\lambda}\TT_{ra}^{(0)}u^a,\nonumber \\
F''&=&\frac{2}{d\lambda^2 G}\TT_{ab}^{(0)}h^{ab}+\frac{1}{4}d(d-1)\frac{FG'^2}{G^2}+\nn\\
&&-\frac{2}{d}(d-1)F\TT_{rr}^{(0)}-\frac{4(d-1)}{d\lambda}\TT^{(0)}_{ra}u^a, \nonumber
\ea
where again the superscript denotes working to zeroth order in $x^a$ derivatives. These equations are
the $(rr)$, $(ra)$ on $u^a$ and $(ab)$ on $h^{ab}$ projection of the Einstein equations, respectively. 
Clearly there are constraints on the bulk stress energy tensor such that the metric takes the required stationary form and these are reflected in the final independent Einstein equation, which implies
\be
{\cal T}^{(0)}_{ar } F \lambda + {\cal T}^{(0)}_{ab} u^{b} = 0. \label{cons1}
\ee
One can understand this latter constraint as follows. Taking the static limit of the metric \eqref{form1}, a bulk stress energy tensor compatible with the symmetries must be characterised by three scalar functions as
\ba
&&{\cal T}^{(0)}_{\mu \nu} dx^{\mu} dx^{\nu}=\\
&&= {\cal T}^{(0)r} dr^2 + {\cal T}^{(0)t} dt \left(dt - \frac{2}{\lambda F} dr\right)
+ {\cal T}^{(0)i} dx^i dx_i;\nn \\
&&= \left ( {\cal T}^{(0)r} - \frac{ {\cal T}^{(0)t}}{\lambda^2 F^2} \right ) dr^2 + {\cal T}^{(0)t} d t_s^2 + {\cal T}^{(0)i} dx^i dx_i,\nn
\ea
where $t_s$ is the Schwarzschild time, such that $dt_s = dt - dr/\lambda F$. Conservation of the bulk stress energy tensor implies that only two out of these three functions are independent. Under a boost the form of the stress energy tensor becomes
\ba
{\cal T}^{(0)}_{\mu \nu} dx^{\mu} dx^{\nu} &=& {\cal T}^{(0)r} dr^2 + \frac{2}{\lambda F} {\cal T}^{(0)t} dr du^a+\\
&&+ {\cal T}^{(0)t} (u_a dx^a)^2 +{\cal T}^{(0)i} h_{ab}dx^{a} dx^b. \nn
\ea
Thus we recover the constraint \eqref{cons1} together with the fact that
\be
{\cal T}^{(0)}_{ab} = {\cal T}^{(0)i} h_{ab} + {\cal T}^{(0)t} u_{a} u_b, 
\ee
i.e. there are only two independent $(ab)$ Einstein equations, that projected onto $h_{ab}$ and that projected onto $u^a u^b$. 

It is useful to write down 
the combination of the Einstein equations which gives the
 Hamiltonian constraint in a radial slicing of the spacetime as 
\be
K^2-K_{ab}K^{ab}=\, ^{d+1}R+2G_{\mu\nu}n^{\mu}n^{\nu},
\ee
where $K_{ab}$ is the extrinsic curvature on the timelike hypersurfaces $\Sigma_c$, $K=K_{ab}\gamma^{ab}$ is its trace, $^{d+1}R$ is the Ricci scalar  of the timelike hypersurface $\Sigma_c$ and $n^{\mu}$ is the unit normal vector to the hypersurfaces $\Sigma_c$. The latter is given
by
\be\label{normalvector}
\vec{n}=\lambda \partial_r+u^a\partial_a ;\quad \underline{n}=\frac{1}{\lambda}dr.
\ee
Since we are interested in the case in which the hypersurface $\Sigma_c$ is intrinsically flat this constraint reduces to
\be
K^2 - K^{ab} K_{ab} = 2\lambda^{-2} {\cal T}^{rr},
\ee
where we use $F(r_c) = 1$. It is straightforward to show that this Hamiltonian constraint
can be rewritten as a constraint equation for the Brown-York stress-energy tensor {\cite{Brown:1992br}} defined on $\Sigma_c$
\be \label{bry}
T_{ab}=2(K\gamma_{ab}-K_{ab})
\ee
as
\be\label{hamiltonian}
d\,T_{ab}T^{ab} - T^2 = - 8\,d\lambda^{-2} {\cal T}^{rr} .
\ee
This constraint effectively defines an equation of state for the fluid. Since the equation is quadratic, one can always find two possible solutions for given data for ${\cal T}^{rr}$ on $\Sigma_c$. In the following section we will interpret the two distinct solutions for generic bulk stress energy tensors.

The $(ar)$ components of the Einstein equations can be expressed as the momentum constraint
\be
D_{c} K^{c}_{a} - D_{a} K = {\cal T}_{a \sigma} n^{\sigma},
\ee
where $D_a$ is the covariant derivative in the induced geometry $\gamma_{ab}$. Requiring that 
the Brown York stress energy tensor is conserved on $\Sigma_c$ implies the constraint 
\be\label{bulkconstraint}
\TT_{ar}\lambda +\TT_{ab}u^b=0, 
\ee
which is indeed satisfied in our equilibrium configurations due to \eqref{cons1}; recall that $F(r_c) = 1$ by construction. 
 
\bigskip

The Brown-York stress energy tensor on the generic timelike hypersurface $\Sigma_c$ can be explicitly expressed in the perfect fluid form as 
\be
T_{ab} = p h_{a b} + \rho u_a u_b
\ee
where $p$ is the pressure and $\rho$ is the energy density, given by
\ba
p &=& \lambda \left ( (d-1) G'(r_c) + F'(r_c) \right );  \label{pressure} \\
\rho &=& - d \lambda G'(r_c)\label{energy}, 
\ea
i.e. the values of the gradients of the metric functions, together with $\lambda$, characterise the pressure and energy density of the fluid. There is an apparent redundancy in these expressions, as the two thermodynamic quantities are expressed in terms of three metric parameters. However, recall that $\lambda$ characterises the rescaling of the time coordinate on the hypersurface $\Sigma_c$, \eqref{redef1}. By choosing the time coordinate to be adapted to this hypersurface one can always take $\lambda =1$ but then the time Killing vector at asymptotic infinity will not have its usual normalisation. Therefore $\lambda$ measures the relative normalisation of the Killing vector.

Now let us move to the interpretation of the fluid on $\Sigma_c$ and let the stress energy tensor of the fluid be denoted $T^{F}_{ab}$. A priori, this cannot be assumed to be precisely the Brown-York stress energy tensor as any linear combination of the latter with 
covariant tensors built from the intrinsic metric $\gamma_{ab}$ on the hypersurface $\Sigma_c$ and its curvature would also be conserved. Given that the induced metric is intrinsically flat, the following stress tensor
\be
T^F_{ab} = C_1 T_{ab} + C_2 \eta_{ab} \label{arb}
\ee
would be conserved for any values of the constants $C_1$ and $C_2$. In the following section we will argue why $C_1$ should be precisely one in a holographic theory, when the induced metric on $\Sigma_c$ is the Minkowski metric. However, as in previous works \cite{Bredberg:2011jq,Compere:2011dx,Compere:2012mt}, $C_2$ remains an ambiguity, which shifts
\be
p \rightarrow p + C_2, \qquad \rho \rightarrow \rho - C_2
\ee
but does not change the combination $(p + \rho)$:
\be
(p + \rho) = \lambda (F'(r_c) - G'(r_c)).
\ee
The latter combination of course appears in the thermodynamic relation
\be
(p + \rho) = s T + \cdots =\lambda F'(r_H) (G(r_H))^{d/2} + \cdots
\ee
where the ellipses denote additional contributions from charges etc and we use
\be\label{entr}
s = 4 \pi (G(r_H))^{d/2};\quad T=\lambda\frac{F'(r_H)}{4\pi}. 
\ee
Therefore $C_2$ does not lead to any ambiguity in the thermodynamic relations.

To understand the role of the prefactor $C_1$
one should consider the effect of a conformal transformation on the hypersurface metric: if one scales $\gamma_{ab} \rightarrow \Omega^2 \gamma_{ab}$, this is equivalent to a rescaling of the coordinates as $x^{a} \rightarrow \Omega x^{a}$. Under such a rescaling, it is easy to see that
\ba
&& p \rightarrow \Omega^{-2} p; \qquad \rho \rightarrow \Omega^{-2} \rho; \\
&& s \rightarrow \Omega^{-d} s; \qquad T  \rightarrow \Omega^{-1} T. \nonumber
\ea
The implication of these scaling relations is as follows. If the thermodynamic relation is satisfied when the induced metric on $\Sigma_c$ is Minkowski, with $T$ being the horizon temperature, then the thermodynamic relation is not satisfied for any non-trivial conformal factor $\Omega$ unless an appropriate prefactor $C_1$ is included.  For the thermodynamic relation to be satisfied, one needs to relate the fluid stress energy tensor $T^F_{ab}$ to the Brown-York stress energy tensor $T_{ab}$ as
\be
T^{F}_{ab} = \Omega^{1-d} T_{ab} + \cdots, \label{conformal}
\ee
where the ellipses denote terms built from the induced metric and, in the general case for which the induced metric is not flat, its curvature. As noted above, this tensor is also conserved, provided that the conformal factor $\Omega$ is independent of the coordinates on the hypersurface. This expression closely resembles the expression for the renormalised stress energy tensor in AdS/CFT and indeed we will discuss the relationship between the two tensors later. 

\bigskip

Finally let us reconsider the alternative forms of the equilibrium metrics \eqref{form2} and \eqref{form3} in the context of this discussion. The background metric for the fluid is the induced metric on $\Sigma_c$, namely
\be
ds^2 = g(r_c) (dx^i dx^i - c_l^2 dt^2) \equiv \gamma_{ab} dx^a dx^b, \label{conf1}
\ee
which is only conformally flat. With this form of the metric the entropy density and temperature are given by
\ba
s &=& 4 \pi (g(r_H))^{d/2} \equiv 4 \pi (G(r_H))^{d/2} (g(r_c))^{d/2}; \\
T &=& \frac{1}{4 \pi c_l} f'(r_H) \equiv \frac{1}{4 \pi} F'(r_H) (g(r_c))^{1/2}. \nn
\ea
The Brown-York tensor associated with a hypersurface $\Sigma_c$ in the metric in \eqref{form2} is
\ba
&&T_{ab} dx^a dx^b =\\
&&=\lambda \left ( (d-1) \frac{g'(r_c)}{g(r_c)}+ \frac{f'(r_c)}{f(r_c)} \right ) g(r_c) \Big(dx^i dx_i+\nn\\
&&- c_l^2 dt^2+ U_a U_b dx^a dx^b \Big) - \lambda d \frac{g'(r_c)}{g(r_c)} (U_a U_b dx^a dx^b). \nn
\ea
which can be rewritten as
\be
T_{ab} = p (\gamma_{ab} + g(r_c) U_a U_b ) + \rho g(r_c) U_a U_b. 
\ee
where the quantities $p$ and $\rho$ are as given in (\ref{pressure}-\ref{energy}). However, these are {\it not} the physical pressure and energy density of the fluid: appropriate conformal factors of $g(r_c)$ must be included in the latter according to the prescription \eqref{conformal} for the thermodynamic relation to be satisfied. 

In equilibrium the choice of coordinate system is merely one of computational convenience but away from equilibrium different choices really correspond to distinct boundary conditions on the hypersurface 
$\Sigma_c$. For example, it is only sensible to keep the induced metric on $\Sigma_c$ fixed as \eqref{conf1} when one extends to the hydrodynamic regime if the conformal factor $g(r_c)$ is independent of the fluid parameters, as well as the speed of light $c_l^2$. If the conformal factor depends on the fluid parameters then implicitly the background metric on $\Sigma_c$ is only flat to leading order in the hydrodynamic expansion. Moreover the fluid stress energy tensor defined in \eqref{conformal} would also only be conserved to leading order in gradients. In practice, as we discuss in the next section, for many bulk stress energy tensors one can immediately integrate the equations of motion to obtain $g(r) = r^2$, which is independent of the fluid parameters.

\section{Examples of fluids} \label{three}

In this section we consider examples of fluids with various bulk stress energy tensors. In particular, we explore the interpretation of the two solutions for the equation of state \eqref{hamiltonian}. 

\subsection{Negative cosmological constant}\label{sec:bbthermodynamics}

Our first example is the case in which the bulk stress energy tensor is a negative cosmological constant, namely ${\cal T}_{\mu \nu} = - \Lambda g_{\mu \nu}$. 
In this case the Hamiltonian constraint (\ref{hamiltonian}) can be written
in terms of the metric \eqref{ansatz} as
\be
d \,\tau^\prime \bigg(f^\prime +(d-1)f \tau ^\prime\bigg)+2\Lambda =0,
\ee
where $e^{2 \tau(r)} = g(r)$. This 
gives two possible equations of state for the fluid
\be\label{eqstate}
\rho=-\frac{p\,d}{(d-1)}\pm\frac{1}{(d-1)}\sqrt{d^2\,p^2+8\,d\,(d-1)\Lambda},
\ee
from which we then obtain two possible geometries for two distinct dual fluids.
Notice that in the
limit $\Lambda\rightarrow 0$ we recover the two already known results, namely the Rindler equation
of state  $\rho=0$, see e.g. \cite{Compere:2012mt}, and the Taub geometry \footnote{The Taub geometry is a vacuum, homogeneous but anisotropic solution of Einstein gravity first found in four bulk spacetime dimensions in \cite{Taub:1951}.There is a curvature singularity at $r=0$ which is timelike and naked.}
 equation of state $\rho=-\frac{2d}{d-1}p$, see \cite{Eling:2012ni}.

The general solution to the $(rr)$ Einstein equation is
\be
g(r)= (c_1 r+ c_2)^2 \qquad\text{with}\qquad c_1\neq 0.
\ee
Here $c_1$ and $c_2$ denote integration constants. The special case $c_1=0$ and $g(r)$ constant is not compatible with the other equations of motion
unless $\Lambda =0$. Let us set $c_1=\frac{1}{L}$,  rescale the $r$ coordinate $\frac{\bar{r}}{L}=\frac{ r}{L}+c_2$ and
use the same symbol for the radial coordinate $\bar{r}\rightarrow r$. The remaining Einstein equations provide a solution for $f(r)$ which reads
\be \label{h}
f(r)=-\frac{2}{d(d+1)}\Lambda r^2+\frac{c_3}{r^{d-1}}.
\ee

Consider first the case $c_3\neq0$.
Inserting into (\ref{h}) the value of the cosmological constant\\ $\Lambda=-\frac{d(d+1)}{2L^2}$ relevant to the maximally symmetric spacetime with negative curvature, the static metric  (\ref{ansatz}) represents a ``naked-brane'' or a black-brane geometry depending on the sign of $c_3$, i.e. the metric is
\be
ds^2=-\frac{r^2}{L^2}\bigg(1+\frac{c_3L^2}{r^{d+1}}\bigg)dt^2+\frac{r^2}{L^2}dx^idx_i+2dtdr.
\ee
{Note that in the limit $\Lambda\rightarrow0$ the Taub geometry with a positive or negative $g_{tt}$ (depending on the sign of $c_3$) is recovered; the limit is obtained by scaling $x^i/L \rightarrow x^i$ with $L \rightarrow \infty$ and $(t,r)$ finite.}
The case with $c_3>0$ is the well-known unphysical negative mass black brane, with naked singularity at $r=0$, and we shall not consider it further here.  Conversely setting $c_3L^2=-r_H^{d+1}$,
the resulting geometry is that of a (positive mass) AdS black brane in ingoing Eddington-Finkelstein coordinates:
\ba\label{bb}
ds^2&=&-\frac{r^2}{L^2}h(r)dt^2+\frac{r^2}{L^2}dx^{i}dx_i+2dtdr, \\
h(r)&=&\left ( 1-\bigg(\frac{r_H}{r}\bigg)^{d+1} \right ). \nonumber
\ea 
The dual fluid living on a generic timelike hypersurface $\Sigma_c$ of constant $r_c$
satisfies the equation of state (\ref{eqstate}) with a positive sign. Boosting and then rescaling to bring the metric into the form \eqref{form1} with
\be\label{GFbb}
G(r)=\frac{r^2}{r_c^2};\quad F(r)=\frac{r^2}{r_c^2}\frac{h(r)}{h_c};\quad \lambda = \frac{r_c}{L}\sqrt{h_c},
\ee
its thermodynamic properties can be  obtained immediately from (\ref{pressure}-\ref{energy}) 
\ba
\rho&=&-\frac{2d}{L} h_c^{1/2} ,\label{rrho}\\
p&=&\frac{1}{L h_{c}^{1/2}} \left ( 2d \,h_c + r_c h'(r_c) \right ),\label{ppp}
\ea
where here and in what follows we use the shorthand notation $h_c \equiv h(r_c)$. 
Notice that in the near horizon limit $r_c\rightarrow r_H$ the pressure and energy density behave as
\ba
\rho&\rightarrow&\OO(r_c-r_H)^{\frac{1}{2}},\label{rhohor}\\
p&\rightarrow &\sqrt{(d+1)}\frac{\sqrt{r_H}}{L}\frac{1}{\sqrt{r_c-r_H}}+\OO(r_c-r_H)^{\frac{1}{2}}\label{phor}
\ea
and thus have the same behaviour as the corresponding thermodynamic quantities of the Rindler fluid\footnote{After restoring the dimensionful parameter $k$, the flat spacetime metric in Rindler coordinates is written $ds^2=-(kr)^2d\hat{t}^2+dx^idx^i+dr^2$ and in ingoing Rindler coordinates is $ds^2=-2krdt^2+dx^idx^i+2dtdr$. It is straightforward to show that the thermodynamic quantities of the dual fluid living on a generic timelike hypersurface $\Sigma_c$ of constant $r=r_c$ are     $p_R=\sqrt{2k}\frac{1}{\sqrt{r_c}}$, $\rho_R=0$.}, see \cite{Compere:2012mt}, provided that the constant $k$ is identified with the surface gravity of the black brane as
\be 
k=\frac{(d+1)}{2L^2}r_H.
\ee 
The energy density (\ref{rrho}) is negative, but as already mentioned one can use the ambiguity in the definition of the Brown-York stress-energy tensor (\ref{arb}) in order to shift the energy density to a positive value. Moreover, the thermodynamic relation 
\be\label{thermrel}
(\rho + p) = \frac{d+1}{L h_c^{1/2}} \frac{r_H^{d+1}}{r_c^{d+1}}
\ee
is independent of this ambiguity. 
The Hawking temperature computed from the metric \eqref{form1} with the time coordinate rescaled is 
\be\label{Hawrescaled}
T_H =\frac{(d+1)}{4\pi r_c L}\frac{r_H}{\sqrt{h_c}}.
\ee
The entropy density of the horizon, again expressed in terms of the rescaled coordinates, is
\be
s = 4 \pi \frac{r_H^{d}}{r_c^d},
\ee
and thus the thermodynamic relation (\ref{thermrel}) is indeed satisfied. The $c_{3} \rightarrow 0$ limit gives 
the AdS vacuum metric expressed in ingoing coordinates
\be\label{metric1}
ds^2=-\frac{r^2}{L^2}dt^2+\frac{r^2}{L^2}dx^idx^i+2dtdr.
\ee
with the energy density and the pressure being
\be	
\rho=-\frac{2d}{L},\qquad p=\frac{2d}{L},
\ee
where again these are evaluated in coordinates in which the induced metric on the hypersurface is Minkowski. 
These values may be viewed as the boundary limit $(r_c\rightarrow\infty)$ and the vacuum AdS limit ($r_H\rightarrow 0$) of the thermodynamic quantities found in (\ref{rrho}-\ref{ppp}).

The arbitrariness in the definition of the fluid stress energy tensor \eqref{arb} can be used to redefine the pressure of the dual fluid in order to obtain a vanishing renormalized ideal stress energy tensor $T^F_{ab}$ for the vacuum, choosing $C_2 = - 2d/L$.  The combination $(\rho + p)  =0$ is invariant and reflects the fact that 
the AdS spacetime metric  (\ref{metric1}) has a trivial horizon geometry and  thus one cannot associate to it a non-zero entropy. One could however define an Unruh temperature
for the observers on timelike hypersurfaces which is non vanishing but fixed
\be
T=\frac{1}{2\pi L}.
\ee
The thermodynamic relation $(p+\rho)=Ts$ is trivially satisfied in this limit since the entropy density $s$ is  zero.

Before turning to the other independent solution of the Hamiltonian constraint, we should note the following. Suppose we instead choose the induced metric on $\Sigma_c$ to be the conformally rescaled metric
\be
ds^2 = \frac{r_c^2}{L^2} ( - d \bar{t}^{\prime 2} + d\bar{x}^{\prime i}  d\bar{x}^{\prime}_{ i}),
\ee
in terms of which coordinates 
\be
s = 4 \pi \frac{r_H^d}{L^d}; \qquad
T_{H} = \frac{(d+1)}{4\pi L^2}\frac{r_H}{\sqrt{h_c}}.
\ee
In order for the thermodynamic relation to be satisfied, then as anticipated we need to use (\ref{conformal}) 
to define the energy density and pressure; they are rescaled by a factor of $r_c^{d+1}/L^{d+1}$ relative to 
(\ref{rrho}) and (\ref{ppp}). 

The effective temperature clearly diverges as $\Sigma_c$ approaches the horizon but this divergence evidently 
arises only from the time coordinate rescaling. In other words, in rescaling time so that the induced metric on $\Sigma_c$ is flat, the effective energy density, pressure and temperature diverge as the hypersurface approaches the horizon. If one worked instead with the coordinate $t$ the temperature and energy density would remain finite but the induced metric (the background metric for the fluid) would as expected become null in this limit. 

\bigskip

Finally let us turn to the interpretation of the equation of state with a negative sign. To understand this it is useful to rewrite the equation of state in the form
\be
\left (1 - \frac{1}{d} \right ) \rho + p = \pm \sqrt{p^2 - 8 \frac{(d-1)}{d} \lambda^{-2}{\cal T}^{rr}},
\ee
which reduces to the equation given in \eqref{eqstate} in the case ${\cal T}^{rr} =- \lambda^2\Lambda$. The solution given above solves this equation with a positive sign. A corresponding solution for the equation of state with a negative sign is obtained by taking
\be
\rho \rightarrow - \rho; \qquad
p \rightarrow - p,
\ee
whenever ${\cal T}^{rr} \neq 0$. (When ${\cal T}^{rr} =0$, i.e. vacuum solutions, the equations of state degenerate. The positive sign above gives $\rho = 0$ while the negative sign gives the Taub equation of state.) Now the switch in signs in the energy density and pressure can be achieved by switching the direction of the normal to the hypersurfaces, i.e. 
\be
n^{\mu} \rightarrow - n^{\mu},
\ee
which corresponds to changing the sign of the extrinsic curvature of the hypersurface.
Physically, however, the negative sign solution gives a negative value for $(\rho + p)$ and therefore the thermodynamic relation $(\rho + p) = s T$ could only be satisfied by a negative temperature. Therefore the second equation of state never gives physically meaningful solutions.

\subsection{Positive cosmological constant}

We now turn to the case in which the bulk stress energy tensor is a positive cosmological constant.
The spacetime metric (\ref{ansatz}) solving  Einstein equations with positive cosmological constant $\Lambda=\frac{d(d+1)}{2l^2}$ corresponding to  $c_3 \ge0$ in (\ref{h})
is 
\be
ds^2=f(r)dt^{2}+ \frac{r^2}{l^2} dx^{i}dx^i+2dtdr,\label{ds}
\ee
with 
\be\label{hhh}
f(r)=\frac{r^2}{l^2} \bigg(1-\bigg(\frac{r_H}{r}\bigg)^{d+1}\bigg).
\ee
This metric is related to the $AdS_{d+2}$-black brane (\ref{bb}) through a Wick rotation of the time coordinate $t$ (i.e. going to Euclidean $AdS$-type coordinates), of the radial coordinate $r$ and of the $AdS$ radius $L$   
\be
t\rightarrow -it;\quad
r\rightarrow ir;\quad
l^2\rightarrow -L^2.
\ee
The induced metric on hypersurfaces of constant $r$ is positive definite and the unit normal vector is now timelike\footnote{The unit normal vector is defined as
\be
\underline{n}=\frac{1}{\sqrt{\pm g^{rr}}}dr,
\ee
with a plus sign if it is spacelike or a minus  sign if it is timelike.
Moreover, there is always an arbitrariness in the redefinition $n_{\mu}\rightarrow -n_{\mu}$
depending on the direction of the vector. As in the previous case we choose the overall sign so that the hypersurface has positive extrinsic curvature. 
}
\be
\underline{n}=-\frac{1}{\sqrt{f}}dr.
\ee
The metric (\ref{ds}) is a de Sitter brane in the so-called inflationary patch; de Sitter is not static so one could not expect the ansatz to produce a spacetime foliated by timelike hypersurfaces. The existence of the analytic continuation to AdS guaranteed that one could describe de Sitter in this ansatz, with the analytic continuations indicated above. 

One can write down the Brown-York tensor on the hypersurface, but it cannot be interpreted
as a perfect fluid stress energy tensor since the hypersurface is spacelike. By construction it can of course be interpreted as the analytic continuation of such a fluid stress energy tensor. 
Note that the metric (\ref{ds}) can be written in the more familiar planar $dS$ coordinates after performing the following change of variables
\be
dt =dy-\frac{dr^2}{h(r)},\qquad
r^2 = \frac{l^4}{\tau^2},
\ee
the resulting  metric being
\be
ds^2=\frac{l^2}{\tau^2}\bigg(h(\tau)dy^{2}+dx^{i}dx^i-\frac{d\tau^2}{h(\tau)}\bigg),
\ee
with
\be
h(\tau)=1-\bigg(\frac{\tau}{\tau_H}\bigg)^{d+1}.
\ee
This is the $dS_{d+2}$-``S brane''  metric\footnote{Referring to this as a brane, while common in the literature, is somewhat of a misnomer since $\tau = \tau_H$ is a coordinate singularity rather than a brane horizon.} in planar coordinates after identifying $l$ with the $dS$ radius and $\tau$ being the $dS$ (inflationary) time coordinate.

\subsection{Fluids for which ${\cal T}_{rr} = 0$}

Certain matter bulk stress energy tensors compatible with the static ansatz (\ref{ansatz}) will in addition satisfy ${\cal T}_{rr}=0$.  The significance is that if ${\cal T}_{rr}$ vanishes, then one can immediately integrate the $(r r)$ Einstein equation to obtain
\be
g(r) = (c_1 r + c_2)^2.
\ee
This integrability is, as mentioned earlier, related to the nested hierarchy found in the Bondi-Sachs \cite{Bondi:1962px,Sachs:1962wk} parameterisation for vacuum spacetimes; similarly the case of negative cosmological constant also satisfies this property.
As discussed above, vanishing $c_1$ is generically only consistent with the other Einstein equations when the bulk stress energy tensor is zero. Integrating the $( a r )$ Einstein equations one obtains the solution for $f(r)$ in terms of the bulk stress tensor component ${\cal T}_{ t r}$
\be
f(r) = \frac{c_3}{( r + c_2/c_1)^{d-1}} + \frac{2}{d} (r + c_2/c_1)^{1-d} \int^r dr' {\cal T}_{r' t} (r' + c_2/c_1)^d.
\ee
In the case where $c_1  =0$ the remaining Einstein equations imply $f'' = 0$. When $c_1$ is non-zero one can absorb both constants into a redefinition of the origin and scale of the radial coordinate, $(c_1 r + c_2) \rightarrow r$ and hence 
\be
f(r) = \frac{c_3}{r^{d-1}} + \frac{2}{d r^{d-1}} \int^r dr' {\cal T}_{r' t} (r')^d.
\ee
The class of stress energy tensors for which ${\cal T}_{rr} =0$ includes in particular gauge fields. Noting that a vector field stress energy tensor is expressed in terms of the field strength $F_{\mu \nu}$ as
\ba
{\cal T}(F)_{\mu \nu} &= & 2 \left(F_{\mu \rho} F_{\nu}^{\;  \rho} - \frac{1}{4} F^{\rho \sigma} F_{\rho \sigma} g_{\mu \nu}\right) +\\
&& \qquad  + m^2 \left(A_{\mu} A_{\nu} - \frac{1}{2} A^{\rho} A_{\rho} g_{\mu \nu}\right), \nonumber 
\ea
then the metric ansatz together with the antisymmetry of $F_{\mu \nu}$ forces the $(rr)$ components in the first line to vanish. The second line involves the mass parameter of the vector field and the vector potential $A_{\mu}$; the $(rr) $ components vanish if $m^2 = 0$ (i.e. it is a gauge field) or $A_r = 0$. However the latter is generically not implied by the symmetries of the equilibrium static solution, which permit a non-zero $F_{tr}(r)$ as the massive vector field equation is
\be
\frac{1}{\sqrt{-g}} \partial_{\mu} (\sqrt{-g} F^{\mu}_{\; \nu}) = m^2 A_{\nu},
\ee
and the $(r)$ component of the left hand side is therefore generically non-zero. The exception is when $m^2 =0$: then $A_r$ is still non-zero in a gauge which is regular at the horizon i.e.
\be
\underline{A} = a(r) \left(dt - \frac{dr}{f(r)}\right) = a(r) d {t}_s
\ee
(with ${t}_s$ again the Schwarzschild time) but $A_r$ does not appear in the stress energy tensor. Note that in the specific case of four spacetime dimensions a constant field strength in the two spatial directions is also consistent with the required symmetries; this case is relevant for discussing gravity duals to magnetohydrodynamics \cite{Zhang:2012uy,Lysov:2013jsa}. 

Consider the case in which the bulk stress energy tensor consists of a cosmological constant and a gauge field. The gauge field equation gives
\be
F_{tr} = \frac{q}{r^d} ,
\ee
with the conserved charge being proportional to $q$ and the general solution for $f(r)$ hence becomes
\be
f(r) = \frac{c_3}{r^{d-1}} - \frac{2}{d (d+1)} \Lambda r^2 + \frac{2 q^2}{d (d-1) r^{2(d-1)}}.
\ee
For negative cosmological constant we therefore recover AdS charged branes, as expected. 

{For $\Lambda = 0$} the solution with $c_3 > 0$ describes what might be called a charged Taub fluid: the metric is not asymptotically flat and has a naked singularity at $r= 0$. For $c_3 < 0$ $f(r)$ is positive for $0 < r < r_H$ and negative for $r > r_H$ where $f(r_H) = 0$. In the inner region hypersurfaces of constant $r$ are timelike, but there is a naked singularity and the region is bounded by a horizon. In the outer region hypersurfaces of constant $r$ are spacelike, and both $t$ and $r$ are null coordinates as $r \rightarrow \infty$. 

One can understand the relationship of the latter solution to the regions inside a Reissner-Nordstrom black hole as follows. Consider four-dimensional black holes (the generalisation to $d > 2$ being straightforward). Start from the metric in ingoing coordinates
\be
ds^2  = - \left(1 - \frac{2M}{R} + \frac{Q^2}{R^2}\right) dv^2 + 2 dv  dR + R^2 d \Omega_2^2.
\ee
Now zoom into the neighbourhood of a point on the two sphere, which without loss of generality can be chosen to be the north pole, by letting $\theta = \ep x$ with $\ep \ll 1$ i.e.
\be
d\theta^2 + \sin^2 \theta d \phi^2 \approx \ep^2 (dx^2 + x^2 d \phi^2).
\ee
In addition scale the radial coordinate such that $r = \ep R$ remains finite and the time coordinate such that
$t = v/\ep$ stays fixed, and also holding fixed 
\be
2 m \equiv 2 M \ep^3; \qquad
q \equiv Q \ep^2.
\ee
Under such rescalings one can see immediately using
\be
R_{\pm} = M \pm \sqrt{M^2 - Q^2}
\ee
that the outer horizon at $R_{+}$ is pushed to infinity (in the $r$ coordinate) while the inner horizon at $R_{-}$ remains at a finite value of $r$. The resulting metric is
\be
ds^2 = - \left(\frac{q^2}{r^2} - \frac{2m}{r}\right) dt^2 + 2 dt dr + r^2 (dx^2 + x^2 d \phi^2),
\ee
which is the $d=2$ case of the metric given above. As discussed above this metric covers the region between an outer horizon, an inner horizon and the singularity. 

{For positive cosmological constant}, the solution with $c_3 > 0$ describes a charged solution with a singularity at $r=0$ and a horizon at a finite value of $r = r_H$, as discussed in the previous section. The hypersurfaces of constant $r$ are only timelike in the region $r < r_{H}$. The solution with $c_3 \le  0$ is more interesting: whilst the behaviour of $f(r)$ at very small $r$ and very large $r$ is unchanged, the function can pass through zero more than once in the intermediate region, corresponding to inner and outer horizons. 

\subsection{Fluids for which ${\cal T}_{rr} \neq 0$}

Many common matter Lagrangians induce stress energy tensors which are compatible with the static ansatz (\ref{ansatz}) but do not satisfy ${\cal T}_{rr} = 0$. In such cases the $(rr)$ Einstein equation does not decouple and one cannot in general immediately solve for $g(r)$ (and hence the other defining functions); the Einstein and matter field equations remain coupled.  

To illustrate this point it is useful to consider a class of Lagrangians which have recently received considerable attention in the context of AdS/CMT: (neutral) scalars coupled to vector fields, so-called Einstein-Maxwell-Dilaton models. Expressing the matter action for a single such scalar $\phi$ coupled to a vector field $A_{\mu}$ as
\ba
S_{m} &=& - \int d^{d+2} x \sqrt{-g} \Big( \frac{1}{2} (\partial \phi)^2 + V(\phi) +\nn\\
&&\qquad\qquad +\frac{1}{4} e^{\alpha \phi} F^2 + \frac{1}{2} e^{\beta \phi} m^2 A^2 \Big),
\ea
with the scalar potential and the parameters $(\alpha, \beta,m^2)$ defining the model, then the equations of motion are known to admit  Lifshitz, hyperscaling violating Lifshitz solutions and other charged dilatonic black holes for various choices of these parameters. The matter stress energy tensor is
\ba
{\cal T}_{\mu \nu} &=& \frac{1}{2} (\partial_{\mu} \phi) (\partial_{\nu} \phi)
- \frac{1}{4} (\partial \phi)^2 g_{\mu \nu} - \frac{1}{2} V(\phi) g_{\mu \nu}+\\
&& \qquad  + \frac{1}{2} e^{\alpha \phi}  (F_{\mu \rho} F_{\nu}^{\;  \rho} - \frac{1}{4} F^{\rho \sigma} F_{\rho \sigma} g_{\mu \nu})+ \nonumber \\
&& \qquad  + \frac{1}{2} m^2  e^{\beta \phi} (A_{\mu} A_{\nu} - \frac{1}{2} A^{\rho} A_{\rho} g_{\mu \nu}),\nonumber 
\ea
and the matter field equations are
\ba
&&\Box \phi = V'(\phi) + \frac{1}{4} \alpha e^{\alpha \phi} F^2 + \frac{1}{2} \beta m^2 e^{\beta \phi} A^2; \\
&&\nabla_{\mu} (e^{\alpha \phi} F^{\mu \nu}) = 2 m^2 e^{\beta \phi} A^{\nu}, \nonumber
\ea
with $\nabla_{\mu}$ the covariant derivative. Consistency with the static, spatially homogeneous ansatz requires \footnote{A magnetic flux for a gauge field is again possible only in four spacetime dimensions.}
\be
\phi = \phi (r); \qquad
\underline{A} = a(r) \left(dt - \frac{dr}{f(r)}\right)
\ee
but then ${\cal T}_{rr} \neq 0$ whenever $\phi(r) \neq 0$ and/or $m^2 a(r) \neq 0$. The metric plus matter is characterised by four functions but the equations of motion are coupled and non-linear so cannot be solved analytically in general. For example, in the case of the pure massive vector (no scalar field) an exact solution at zero temperature with Lifshitz scaling symmetry is known, see e.g. \cite{Taylor:2008tg}, but corresponding finite temperature blackened solutions have only been found numerically, see for example \cite{Danielsson:2009gi,Bertoldi:2009vn,Bertoldi:2009dt,Bertoldi:2011zr}. The zero temperature Lifshitz solution can be written in our coordinate system as
\ba
f(r) &=& \frac{r^2}{z^2}; \qquad g(r) = \left ( \frac{r}{z} \right )^{2/z}, \\
a(r) &=& 2 \frac{\sqrt{z-1}}{z^2} r; \qquad
m^2 = \frac{d^2}{2 z^2}. \nn
\ea
In this case the $(rr)$ Einstein equation can be integrated to give an analytic solution for the function $g(r)$, but the latter is no longer given by $g(r) \propto r^2$. The usual form of the Lifshitz metric, i.e.
\be
ds^2 = \frac{d\rho^2}{\rho^2} + \left ( - \frac{d \tau^2}{\rho^{2z}} + \frac{dx^i dx_i}{\rho^2} \right ),
\ee
is obtained by the redefinitions 
\be
r = z \rho^{-z}; \qquad d \tau = \left(dt - \frac{d\rho}{\rho}\right).
\ee

\section{Renormalized versus fluid stress energy tensors} \label{four}

In this section we will explore the relationship between the fluid stress energy tensor which we have defined and the renormalised holographic stress energy tensor \cite{Henningson:1998gx,Balasubramanian:1999re,deHaro:2000xn}. We begin  with a brief discussion of holographic renormalisation for asymptotically locally AdS spacetimes, see \cite{Skenderis:2002wp} for a review.  
Asymptotically one can always express such spacetimes in Fefferman-Graham coordinates 
as
\be
ds^2=L^2 \left ( \frac{d\rho^2}{\rho^2}+\frac{1}{\rho^2}g_{ab}(\rho,x)dx^adx^b \right ) 
\ee
where the metric $g_{ab}$ admits an expansion
\be
g_{ab} (\rho,x) = g_{(0) ab}(\rho,x) + \cdots + g_{(d+1) ab} (\rho,x) \rho^{d+1} + \cdots
\ee
The ellipses denote terms which are subleading as $\rho \rightarrow 0$; the form of the expansion depends on the details of the bulk stress energy tensor but in the case of pure cosmological constant $(g_{(0)},g_{(d+1)})$ are the non-normalizable and normalizable modes, respectively.

Using the defining relation of AdS/CFT {\cite{Gubser:1998bc,Witten:1998qj}}, the expectation value of the dual CFT stress energy tensor is defined as 
\be
\langle T^{\rm CFT}_{ab} \rangle = -\frac{2}{\sqrt{-g_{(0)}}} \frac{\delta S_{\rm ren}}{\delta g_{(0)}^{ab}},
\ee
where we work in Lorentzian signature and $S_{\rm ren}$ denotes the renormalised bulk action, in which counterterms have been added to remove the volume divergences. Algorithmically one expresses this in terms of the induced metric
$\gamma_{ab}$ on a hypersurface of constant $\rho = \rho_c$ and the subtracted bulk action $S_{\rm sub}$ as
\be
\langle T^{\rm CFT}_{ab} \rangle = - \lim_{\rho_c \rightarrow 0} \left [ 
\frac{L^{d-1} }{\rho_c^{d-1} } \frac{2}{\sqrt{-\gamma}}  \frac{\delta S_{\rm sub}}{\delta \gamma^{ab}} \right ],
\ee
where the renormalised action is defined as 
\be
S_{\rm ren} = \lim_{\rho_c \rightarrow 0} S_{\rm sub},
\ee
with the subtracted action being the bare regulated action plus counterterms which remove volume divergent terms. In the specific case of pure cosmological constant the dual stress energy tensor can be expressed as
\be \label{TCFTb2}
\langle T_{ab}^{\rm CFT} \rangle  =\lim_{\rho_c \rightarrow 0} \left [ \bigg(\frac{L}{\rho_c}\bigg)^{d-1}T_{ab}^{\gamma} \right ] = \frac{(d+1)}{L} g_{(d+1)ab} + \cdots,
\ee
where following  \cite{deHaro:2000xn} (and setting $16 \pi G_N = 1$)
\ba
T_{ab}^{\gamma} &=&T_{ab}^{\text{BY}} +T_{ab}^{\rm{ct}},\\
T_{ab}^{\text{BY}}  &=& 2\bigg(K\gamma_{ab}-K_{ab}\bigg), \qquad
T_{ab}^{ {\rm {ct}}}  =- \frac{2 d}{L} \gamma_{ab} + \cdots, \nn
\ea
$T_{ab}^{\text{BY}}$ being the Brown-York tensor and the ellipses in the last expression denote contributions due to additional counterterms  which vanish when the metric  $\gamma$ is flat. Restricting to this case, in which the corresponding boundary metric $g_{(0)}$ is flat, the renormalised CFT stress energy tensor is
\be\label{gdplusone}
\langle T_{ab}^{\rm CFT} \rangle  = \frac{(d+1)}{L} g_{(d+1)ab}.
\ee
One can trivially rewrite \eqref{TCFTb2} as
\be
\langle T_{ab}^{\rm CFT} \rangle  ={\rm{lim}}_{\Omega \rightarrow \infty} \left [ \Omega^{d -1} (T_{ab}^{\rm BY} + C_2 \gamma_{ab} + \cdots)  \right ]\label{adscftprescription}
\ee
where the induced metric on the hypersurface of constant radius is $\Omega^2 \gamma_{ab}$ and $C_2 = -2d/L$. As anticipated, this is very closely related to the expression used for the fluid stress energy tensor but the CFT stress energy tensor is defined using different hypersurfaces. Starting from the generic fluid metric in ingoing coordinates \eqref{form1} one can define Schwarzschild coordinates $x_s^a$ such that
\be \label{coor1}
dx^a = d x^a_s + \frac{u^a dr}{\lambda F(r)},
\ee
so that
\be
ds^2 = \frac{dr^2}{ \lambda^2 F(r)} + G(r) \eta_{ab} dx_s^a dx_s^b + (G(r) - F(r)) u_a u_b dx^a_s  dx^b_s. 
\ee
By defining
\be \label{coor2}
L^2 \frac{d \rho^2}{\rho^2} \equiv \frac{dr^2}{\lambda^2 F(r)},
\ee
the metric can be brought into Fefferman-Graham form. 

For example, in the AdS black brane (\ref{bb}), performing the following change of radial coordinates
\be
r=\frac{L^2}{\rho}\bigg(1+\frac{1}{4}\bigg(\frac{r_H \rho}{L^2}\bigg)^{d+1}\bigg)^{\frac{2}{d+1}},
\ee
with inverse transformation
\be\label{relrhor}
\rho=\frac{L^2}{r}\left(\frac{2}{1+\sqrt{h(r)}}\right)^{\frac{2}{d+1}},
\ee
the black brane metric can be written uniquely in the Fefferman-Graham form  requiring  the boundary  to be at $\rho=0$ and the Dirichlet condition for the representative of the conformal structure at the boundary $g_{(0)ab}=\eta_{ab}$. The transformed metric is
\ba\label{FGexpansion2}
ds^2&=&\frac{L^2}{\rho^2}d\rho^2+\frac{L^2}{\rho^2} g_{\rm fg} (\rho) \eta_{ab}\,dx_s^a dx_s^b+\\
&&+ \frac{L^2}{\rho^2} (g_{\rm fg}(\rho) - f_{\rm fg}(\rho)) (u_a dx_s^a)^2,\nn\\
f_{\rm fg}(\rho) &=& \bigg(1+\frac{1}{4}\bigg(\frac{r_H \rho}{L^2}\bigg)^{d+1}\bigg)^{\frac{2(1-d)}{(d+1)}}\bigg(1-\frac{1}{4}\bigg(\frac{r_H \rho}{L^2}\bigg)^{d+1}\bigg)^2, \nn\\
g_{\rm fg}(\rho) &=& \bigg(1+\frac{1}{4}\bigg(\frac{r_H \rho}{L^2}\bigg)^{d+1}\bigg)^{\frac{4}{d+1}}. \nn
\ea
Notice that the coordinate transformation is singular at the horizon  $\rho_H=2^{\frac{2}{d+1}}L^2/r_H$. A hypersurface of constant $r_c$ is mapped to a radial hypersurface $\rho (r_c)$.  The coordinates are chosen such that the induced metric on a radial hypersurface as $\rho \rightarrow 0$ approaches $ \eta_{ab}\,L^2/\rho^2$. 

Expanding the metric (\ref{FGexpansion2}) around $\rho=0$ one can extract the holographic CFT stress-energy tensor using (\ref{gdplusone}), which is indeed that of a conformal ideal fluid in $(d+1)$ spacetime dimensions
\be
\langle T_{ab}^{CFT}\rangle=  \frac{(4\pi)^{d+1}L^d}{(d+1)^{d+1}} \,T^{d+1}\bigg(h_{ab}+d\,u_au_b   \bigg), 
\ee
where $h_{ab}=\eta_{ab}+u_au_b$. The energy density $\rho_e$, the pressure $p$ and the temperature $T$ of the dual fluid are
\be\label{confp}
\rho_e= \frac{d}{L}\bigg(\frac{r_H}{L}\bigg)^{d+1},\quad p= \frac{1}{L}\bigg(\frac{r_H}{L}\bigg)^{d+1},\quad  T=\frac{(d+1) r_H}{4\pi L^2},
\ee
with the temperature being the Hawking temperature. 

\subsection{Cut-off stress energy tensor}\label{cutoff}

To compare with the discussion in previous sections
we now define the stress energy tensor of a dual fluid living on a generic timelike hypersurface $\Sigma_c$  by taking the expression (\ref{TCFTb2}) on a finite radial coordinate $\rho_c$ 
without taking the limit towards the boundary. In other words, one defines
\be 
\langle T_{ab}^{F} \rangle_{\Sigma_c}   =  \left [ \bigg(\frac{L}{\rho(r_c)} \bigg)^{d-1}T_{ab}^{\gamma} \right ]_{\Sigma_c}. 
\ee
Since the induced metric on this hypersurface is no longer conformally flat, one needs to rescale the $x^a_s$ coordinates as in the previous section. The stress energy tensor takes the form of an ideal fluid with energy density $\rho^F$ and pressure $p^F$
\ba\label{prho2}
\rho^F&=&\frac{d}{L}\bigg(\frac{r_{H}}{L}\bigg)^{d+1}\bigg(1+\frac{A}{4}\bigg)^{-1},\\
p^{F}&=&\frac{1}{L}\bigg(\frac{r_{H}}{L}\bigg)^{d+1}\frac{\bigg( 1+d\frac{A}{4} \bigg)}
{\bigg(1-\frac{A^2}{16}  \bigg)}, \nn
\ea
where $A = (\rho_c r_H/L^2)^{d+1}$. In the limit $r_c\rightarrow\infty$ ($A\rightarrow 0$) one recovers the values given above (\ref{confp}) and the stress-energy tensor is traceless on the boundary as it should be.
Using the relationship (\ref{relrhor}) between $r$ and $\rho$ coordinates\footnote{Notice that  the relation (\ref{relrhor}) is the same as one obtains starting from (\ref{bb}) in rescaled coordinates (\ref{form1}) with (\ref{GFbb}) due to the cancellation of factors in the denominator of the right hand side of (\ref{coor2}). }, (\ref{prho2}) can be immediately rewritten as
\ba\label{qftp1}
\rho^F &=& \frac{2d}{L}\left(\frac{L}{\rho(r_c)}\right)^{d+1} (1 - h_c^{1/2}), \\
p^F &=& \frac{1}{L}\left(\frac{L}{\rho(r_c)}\right)^{d+1}  \left ( 2d (h_c^{1/2} - 1) + \frac{r_c h'(r_c)}{h_c^{1/2} } \right ), \nn
\ea
where
\be
h_c = \left  (1 - \frac{r_H^{d+1}}{r_c^{d+1}} \right );
\ee
which agree with \eqref{rrho} and \eqref{ppp} after  taking into account the inclusion of the counterterm and the conformal factor (in Fefferman-Graham coordinates)  which gives
\ba
\rho^F &=&\left(\frac{L}{\rho(r_c)}\right)^{d+1} \left(\rho + \frac{2d}{L}\right),\\
p^F &=& \left(\frac{L}{\rho(r_c)}\right)^{d+1}\left(p - \frac{2d}{L}\right).
\ea
The usual conformal fluid pressure and energy density (\ref{confp})  can be now obtained after  pushing the cutoff towards the boundary $r_c\rightarrow\infty$.

The agreement between the expressions follows from the form of the coordinate transformations \eqref{coor1} and \eqref{coor2}: restricted to a surface of constant $r$ the ingoing and Schwarzschild coordinates coincide. When one extends the solutions into the hydrodynamic regime, however, the corresponding coordinate transformations will be of the form
\ba
&& dx^a - u^a(x) \frac{dr}{\lambda(x) F(r,x)}  + \cdots = dx^a_s; \\
&& L \frac{d\rho}{\rho} = - \frac{dr}{\lambda(x) F(r,x)} + \cdots, \nn
\ea
where the ellipses are subleading terms in the hydrodynamic expansion. As we will see in the following section, hypersurfaces of constant $\rho$
no longer coincide with hypersurfaces of constant $r$ in the hydrodynamic regime.

\subsection{Interpretation and dual field theory}

\begin{figure}[t] 
\begin{center}
\includegraphics[width=4cm]{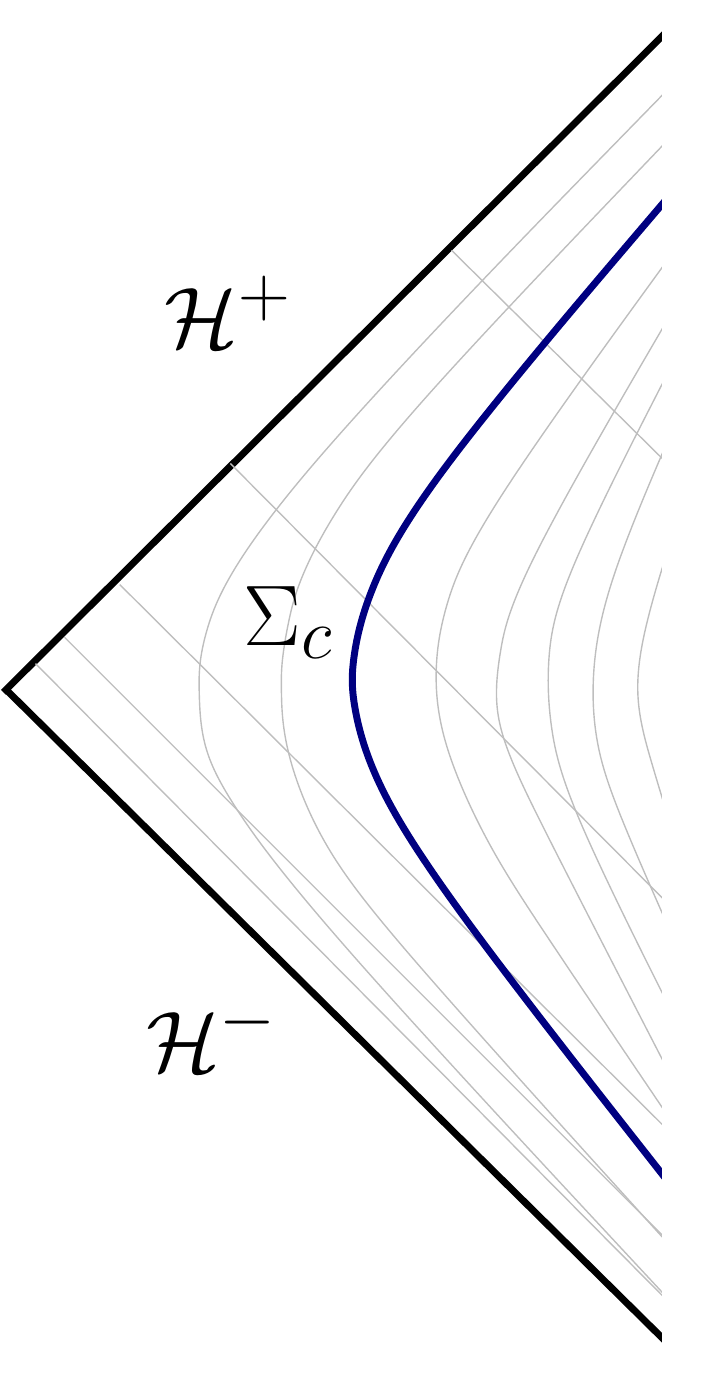}\hspace{1cm}
\begin{minipage}{0.5\textwidth}
\caption{The spacetime has past and future horizons ${\cal H}^-$ and ${\cal H}^+$ respectively.  The dual fluid lives on a timelike surface  $\Sigma_c$.  Lines of constant $t$ and constant $r$ in the Eddington-Finkelstein coordinate system are shown.} \label{fig:Rindler}
\end{minipage}
\end{center}
\end{figure}

In this section we turn to the interpretation in terms of renormalisation group flow of the putative dual field theory. Let us step back from the specific examples to the generic metric ansatz \eqref{ansatz} in which the induced metric $\gamma(r_c)$  on $\Sigma_c$ is given in \eqref{induced}.  The spacetime Penrose diagram is illustrated in Figure \ref{fig:Rindler}. 

In the putative fluid/gravity correspondence, the background metric for the fluid (and hence for the field theory) is $\gamma(r_c)$. Viewing $r_c$ as an energy scale, it is manifest that this metric runs with the energy scale. In field theory language, suppose the field theory has couplings $g^a(\Lambda)$ which run with the energy scale
$\Lambda$ via their beta functions
\be
\beta_{g^a} = \frac{ \partial g^a(\Lambda)}{\partial \ln \Lambda}
\ee
then it is usual to characterise the couplings in terms of a reference (cutoff) energy scale $\Lambda^o$ such that
$g^a(\Lambda^o) \equiv g^a_o$. In the case at hand, the induced metric for the fluid effectively provides background couplings for the putative dual theory. The boundary condition on $\Sigma_c$ imposes 
\be
\gamma(r_c)_{ab} = \Omega^2 \eta_{ab}
\ee
i.e. at this ``energy" scale the metric is conformally flat. Having fixed the metric at this reference point, the metric at a different energy scale is generically not conformally flat since the time and space components ``run" differently. Just as in field theory, however, the choice of reference point is arbitrary: one can change $r_c \rightarrow r_c'$. 

One can only provide a fluid/gravity interpretation if the fluid stress energy tensor is conserved\footnote{Strictly speaking the fluid stress energy tensor should satisfy an appropriate conservation equation but is not always conserved. The generalisation to cases in which it is not conserved because of e.g. sources for currents is straightforward.} and satisfies a thermodynamic relation with the temperature and entropy density. 
As many authors have pointed out, the Brown-York stress energy tensor defined in \eqref{bry}
is a natural candidate for the fluid energy momentum tensor as it is conserved. One has to be careful here, though, since a fluid stress energy tensor defined as 
\be\label{fluidstress}
T^F_{ab} = C_1(r_c) T_{ab}^{\rm BY} + C_2(r_c) \eta_{ab}
\ee
will be conserved for any choice of $(C_1(r_c), C_2(r_c))$. The function $C_2(r_c)$ drops out of the combination
$(\rho + p)$ appearing in the thermodynamic relation but $C_1(r_c)$ is not restricted by conservation. We argued
previously that $C_1(r_c)$ must scale homogeneously under a conformal rescaling of the induced metric so that the thermodynamic relation is satisfied for any choice of the conformal factor: $C_1(r_c) \rightarrow \Omega^{1-d} C_1(r_c)$ as $\eta_{ab} \rightarrow \Omega^2 \eta_{ab}$. 

Clearly the thermodynamic relation is only satisfied for a specific choice of $C_1(r_c)$ (for fixed induced metric 
$\eta_{ab}$).  The choice of $C_1(r_c)$ defines the holographic dictionary for the fluid. For $\Omega = 1$ the choice of $C_1(r_c)$ which is consistent with the thermodynamic relation is precisely $C_1(r_c) = 1$, in both the Rindler and the AdS cases. We will now show why this must be the case in {\it any} putative holographic correspondences (in which the bulk spacetime is describable by Einstein gravity with matter) satisfying a GKPW  type dictionary  \cite{Gubser:1998bc, Witten:1998qj}, i.e. a duality between the onshell bulk action with ``boundary condition" $\gamma(r_c)$
and the dual generating functional $W[\gamma(r_c)]$. We do not need to assume a negative cosmological constant or indeed any specific form for the bulk stress energy tensor. 

The proof makes use of a Hamiltonian description of the bulk dynamics, following the same approach as in the Hamiltonian method of holographic renormalisation \cite{Papadimitriou:2004ap}. One difference relative to the latter is that we will work with a finite cutoff and we will not need to look in detail at the renormalisation. A second difference relative to Hamiltonian holographic renormalisation is that we do not consider generic bulk solutions with given asymptotics; instead we restrict to the hydrodynamic regime in which the bulk solution is near to an equilibrium solution with horizon. 

Let us consider a $(d+2)$ dimensional manifold which can be radially foliated by hypersurfaces $\Sigma_{r}$ of constant $r$. The metric can then be decomposed as
\be
ds^2 = (N^2 + N_{a} N^{a}) d r^2 + 2 N_{a} dx^a d r + \gamma_{ab} dx^a dx^b,
\ee
with $N$ the shift and $N_a$ the lapse. This metric reduces to the form used in the previous sections upon fixing the gauge such that $(N^2 +N^a N_a) = 0$, which is always possible since $N_a$ is timelike; comparing with the (\ref{form1}), one sees that
\be
N_a = - \frac{u_a}{\lambda}; \qquad
\gamma_{ab} = G(r) \eta_{ab} + (G(r) - F(r)) u_a u_b. 
\ee
However, one cannot not fix this gauge before deriving the constraint equations. 
The action for Einstein gravity coupled to an arbitrary matter Lagrangian ${\cal L}_m$ (which may include a cosmological constant) can be written as 
\be \label{ham-f}
S = - \frac{1}{2 k^2_{(d+2)}} \int d^{d+2}x \sqrt{-\gamma} N \left ( \hat{R} + K^2 - K_{ab} K^{ab} - {\cal L}_m \right )
\ee
where $\hat{R}$ is the curvature of the metric $\gamma_{ab}$ and $K_{ab}$ is the extrinsic curvature of the hypersurface  $\Sigma_{r}$. Note that in this rewriting of the bulk Einstein-Hilbert action one obtains a boundary term which precisely cancels the Gibbons-Hawking term; the variational problem for this action is well defined for given boundary data $\gamma_{ab}$. The canonical momentum conjugate to $\gamma_{ab}$ is 
\be
\pi^{ab}  = \frac{\delta L}{\delta (\partial_{r} \gamma^{ab})} = - \frac{\sqrt{-\gamma}}{2 k_{(d+2)}^2} (K \gamma^{ab} - K^{ab})
\ee 
where the Lagrangian $L$ is defined via $S = \int d r L$. Note that the momenta conjugate to the shift and lapse vanish, implying that the equations of motion are constraints. The crucial next step is to use the Hamilton-Jacobi formalism of mechanics to express the momenta on any given hypersurface as variations of the on shell action with respect to the induced values of the fields on this surface, namely
\be
\pi^{ab} (r_c) = \frac{\delta S_{\rm onshell}} {\delta \gamma_{ab} (r_c)}.
\ee
Note that $r_c$ is arbitrary, with the relation holding for any $r_c$ provided that the radial coordinate is well-defined. If a GKPW dictionary of the form given in {\cite{Gubser:1998bc,Witten:1998qj}} holds, then the onshell action acts as the generating functional for background metric $\gamma_{ab}$ and the corresponding 
stress energy tensor is defined as
\be
T^{ ab} = - \frac{2}{\sqrt{-\gamma}} \frac{\delta S_{\rm onshell}} {\delta \gamma_{ab} (r_c)}.
\ee
Using the Hamilton-Jacobi relation this quantity becomes the Brown-York tensor, with constant prefactor
\be
T^{ ab} = \frac{1}{k_{(d+2)}^2} (K \gamma^{ab} - K^{ab}).
\ee
Therefore the identification of the onshell action with the generating functional of the dual theory in a background metric $\gamma_{ab}$ is equivalent to the stress energy tensor of the dual theory being the Brown-York tensor. 

The only assumption so far is the existence of a radial foliation. If one further insists that such a foliation is by timelike hypersurfaces then it would be valid outside a horizon but clearly not applicable inside a horizon. Moreover, whilst the variational problem is well-defined for the action \eqref{ham-f} given the metric on the bounding hypersurface, the variational problem would be equally well posed if one added to the action boundary terms $S_{B}[\gamma] $ depending only on quantities intrinsic to the induced geometry. The corresponding stress energy tensor would then become
\be
T^{ab} \rightarrow T^{ab} - \frac{2}{\sqrt{-\gamma}} \frac{ \delta S_{B}[\gamma] }{\delta \gamma_{ab}}.
\ee
The boundary terms are uniquely determined in AdS/CFT when one takes the conformal class of the metric to be fixed as one takes the boundary to infinity and these terms are equivalent to the counterterms needed to obtain finite renormalised quantities. At finite $r_c$ there is no natural way to fix the ambiguity with a generic bulk solution but, as we saw earlier, if we restrict to the hydrodynamic regime with the induced metric being flat, the ambiguity does not affect the thermodynamic relation. If the boundary metric for the dual theory is $\tilde{\gamma}_{ab} = \Omega^2 \gamma_{ab}$ then
\be
T^{\tilde{\gamma}}_{ab} \equiv   -\frac{2}{\sqrt{-\tilde{\gamma}}} \frac{\delta S_{\rm onshell}} {\delta \tilde{\gamma}^{ab} (r_c)} = \Omega^{1-d} T^{\gamma}_{ab}\label{prescription},
\ee
in accordance with the relation noted earlier.

\section{Hydrodynamics and near equilibrium solutions} \label{five}

In this section we will promote general seed equilibrium solutions  to hydrodynamic solutions by allowing the fluid properties to become slowly varying. Following the discussion earlier, we consider the form (\ref{form1}) of the seed metric which we rewrite below for convenience:
\be
ds^2 = - \frac{2}{\lambda} u_a d {x}^a dr + {G}(r) d {x}_a d {x}^a + \left ( {G}(r) - {F}(r) \right ) (u_a d {x}^a)^2, \label{form1a}
\ee
in which $G(r_c) = F(r_c) =1$, while $F(r_H) = 0$ at the horizon ${\cal H}$. 

In equilibrium one can always find a shifting and subsequent rescaling of the radial coordinate to set $\lambda =1$.
The flat spacetime metric in static Rindler coordinates  can for example  be written in the form (\ref{form1a}) with $\lambda = 1$ as
\be
ds^2 = 2 dr dt - ( 1 + p ( r - r_c)) dt^2 + dx^{i} dx_{i}.
\ee
The identification of the pressure and energy density of the dual fluid to the Rindler geometry of 
\cite{Compere:2012mt} follows from (\ref{pressure}-\ref{energy}) straightforwardly: the pressure is indeed the parameter $p$ while the energy density is zero.

Scaling $\lambda$ to one is generically not compatible with imposing that the induced metric on the cutoff hypersurface is Minkowski, however, and we will find it more convenient to retain a generic $\lambda$ in what follows. For example, for the AdS black brane geometry (\ref{bb}) the seed metric is conveniently expressed after rescaling of the field theory coordinates as 
\be\label{blackbranemetric}
ds^2 = \frac{2 dt dr}{ \sqrt{h_c}\,r_c/L} - \frac{r^2 h(r)}{r_c^2 h_c} dt^2 + \frac{r^2}{r_c^2} dx^i dx_i,
\ee
in which we again define $h_c = h(r_c)$. 

\subsection{General hydrodynamic equations}

In order to move to the hydrodynamic regime, one needs to promote the thermodynamic parameters to become slowly varying functions, see \cite{Bhattacharyya:2008jc}. This can be achieved by writing 
\ba
ds^2 &&= -2 \lambda (x)^{-1} u_a(x) dx^a dr  + \\
&&+\Big(G(r,x)) \eta_{ab}\! +\! (G(r,x)\! -\! F(r,x))u_a(x)u_b(x)\Big) dx^adx^b,\nn 
\ea
such that
\ba
\partial_a u^b(x)\sim {\cal O}(\epsilon); &\qquad&
\partial \lambda \sim {\cal O}(\epsilon);\\
\partial_a F(r,x) \sim {\cal O}(\epsilon); &\qquad&
\partial_a G(r,x) \sim {\cal O}(\epsilon),\nn
\ea
with $\epsilon$ small. The inverse metric is given by
\ba
g^{rr} &=& \lambda(x)^2 F(r,x); \qquad g^{ra} = \lambda(x) u^a(x); \\
g^{ab} &=& (G(r,x))^{-1} (\eta^{ab} + u^a u^b) \equiv (G(r,x))^{-1} h^{ab}. \nonumber
\ea
By construction this generalisation preserves the induced metric on $\Sigma_c$ and satisfies the Einstein plus matter equations to leading order in $\epsilon$ with the fluid parameters as given above. To any order $n \ge  1$ one will need to correct the metric by terms $g^{(n)}_{\mu \nu}$ such that the Einstein equations are still satisfied. Conservation of the Brown York stress tensor on $\Sigma_c$ to leading order in gradients implies
\ba
(p + \rho) D^a u_a &=& - D \rho; \label{cons} \\
(p + \rho) a_c &=& - D_c^{\perp} p, \nonumber
\ea
where $D_a^{\perp} \equiv h^{b}_{a} \partial_b$,  $D \equiv u^a \partial_a$ and the acceleration $a_c = D u_c$. Note that the Rindler fluid had the property that the equilibrium energy density vanishes, and thus the fluid was incompressible to leading order, but this property is not in general satisfied. 

Supposing that one weights derivatives such that $\partial_r \sim 1$ and $\partial_a \sim \epsilon$, if one adds a piece $g^{(n)}_{\mu \nu}$ to the metric at order $\epsilon^n$ then one can compute the associated change in the Ricci tensor at order $\epsilon^n$ to be $\delta R^{(n)}_{\mu \nu}$. This is computed using the usual linearised formula
\ba
\delta R^{(n)}_{\mu \nu} &=& \frac{1}{2} \Big( \nabla^{\lambda} \nabla_{\mu} g^{(n)}_{\nu \lambda}
+ \nabla^{\lambda} \nabla_{\nu} g^{(n)}_{\mu \lambda}+\\
&&- \nabla^{\lambda} \nabla_{\lambda} g^{(n)}_{\mu \nu}
- \nabla_{\mu} \nabla_{\nu} {\rm tr} (g^{(n)})
\Big),\nn
\ea
with $\nabla$ evaluated using the leading order metric given above. Note that, since one needs to retain terms in
the covariant derivative at order $\ep^0$, only radial derivatives need to be retained.

Order by order we must require the additional terms in the metric to be such that the Einstein equations are satisfied. Defining $\hat{R}_{\mu\nu}^{(n)}$ as the part of the  Ricci tensor corresponding to  $g^{(n-1)}$ due to  $\partial_a$ derivatives and $\delta R_{\mu\nu}^{(n)}$ being the part of the Ricci tensor related to radial derivatives $\partial_r$ of $g^{(n)}$, the Einstein equations at order $n$ are
\be\label{condition1}
\hat{R}_{\mu\nu}^{(n)}+\delta R_{\mu\nu}^{(n)}-\frac{1}{2}\sum_{k=0}^{n}g_{\mu\nu}^{(k)}\hat{R}^{(n-k)}-\frac{1}{2}g_{\mu\nu}^{(0)}\delta R^{(n)} =\TT_{\mu\nu}^{(n)},
\ee
where the Ricci scalar at each order is given by 
\be
\hat{R}^{(n)}+\delta R^{(n)}=\sum_{k=0}^n \hat{R}_{\alpha\beta}^{(k)}g^{(n-k)\alpha\beta}+\delta R^{(n)}_{\alpha\beta}g^{(0)\alpha\beta},
\ee
and the inverse metric up to order $k$ is defined in the way to assure that the trace of the metric is always the same ${\rm tr}(g)=d+2$, for example at first order it is
\be
g^{(1)\mu\nu}=-g^{(0)\mu\rho}g^{(0)\nu\sigma}g_{\rho\sigma}^{(1)}.
\ee
Conditions (\ref{condition1}) can be then rewritten in a more compact way as
\ba\label{integrability}
&&\hat{R}_{\mu\nu}^{(n)}+\delta R_{\mu\nu}^{(n)}=\bar{\TT}_{\mu\nu}^{(n)}, \\
&&\bar{\TT}_{\mu\nu}^{(n)}=\TT_{\mu\nu}^{(n)}-\frac{1}{d}\sum_{k=0}^ng_{\mu\nu}^{(k)}\TT^{(n-k)}, \nn
\ea
with
\be
\TT^{(n)}=\sum_{k=0}^{n}\TT_{\alpha\beta}^{(k)}g^{(n-k)\alpha\beta}.
\ee

Imposing the natural gauge choice $g^{(n)}_{r \mu} = 0$ for $n \ge 1$  we obtain the following for the perturbations of the Ricci tensor induced by the addition of $g^{(n)}$:
\ba\label{Riccivar}
\delta R^{(n)}_{rr} &=& - \frac{1}{2} h^{ab} \partial_r^2 \left ( \frac{1}{G} g^{(n)}_{ab} \right )- \frac{1}{2} \frac{G'}{G}h^{ab} \partial_r \left ( \frac{1}{G} g^{(n)}_{ab} \right );\nn \\
\lambda ^{-1} \delta R^{(n)}_{ra} &=&\frac{1}{2G^{d/2}} u^{b} \partial_r\left ( G^{d/2}\partial_r g^{(n)}_{ab}\right)+\nn\\
&&+ \frac{1}{4} F' u_a 
h^{cd} \partial_r \left ( \frac{1}{G} g^{(n)}_{cd} \right )+\nn\\
&&
-\frac{1}{2G^{d/2}}u^c h^d_a\partial_r\left( \frac{G'}{G}G^{d/2}g^{(n)}_{cd}\right);  \\
\lambda^{-2} \delta R^{(n)}_{ab} &=& - \frac{1}{4} F \left(G' h_{ab} -F' u_au_b\right) h^{cd} \partial_r \left( \frac{1}{G} g^{(n)}_{cd}\right)+\nn\\
&&
- \frac{1}{2 G^{d/2}} \partial_r \left( G^{d/2} F \partial_r g^{(n)}_{ab}\right)+\frac{G'F}{G}h^c_{(a}\partial_rg^{(n)}_{cb)}+ \nonumber \\ && +\frac{F'G'}{G}u_{(a}h^c_{b)}u^dg^{(n)}_{cd}-F'u_{(a}u^c\partial_rg^{(n)}_{cb)}+\nn\\
&&-\frac{1}{2}\frac{FG'^2}{G^2}h^c_{(a}h^d_{b)}g^{(n)}_{cd}-\frac{1}{2}\frac{G'^2}{G}h_{ab}u^cu^dg^{(n)}_{cd}+ \nonumber \\
&& +
\frac{1}{2 G^{d/2}}u^cu^d\partial_r\left( G^{d/2}(G'h_{ab}-F'u_au_b)g^{(n)}_{cd}\right). \nonumber 
\ea
where primes denote radial derivatives.
Moreover we find
\ba\label{generaleq}
\lambda^{-2}\delta R^{(n)}&=&-\frac{1}{2}\frac{F}{G^{d/2}}\partial_r\left( G^{d/2}\partial_r\left(\frac{1}{G}g_{cd}^{(n)}h^{cd}\right)\right)+\nn\\
&&
-\frac{1}{2}F^{\prime}\partial _r\left( \frac{1}{G}g_{cd}^{(n)}h^{cd}\right)+\frac{G^{\prime}F}{2G^2}\partial _rg_{cd}^{(n)}h^{cd}+\nn \\
&&-\frac{1}{2G^{d/2+1}}\partial_r\left( G^{d/2}F\partial _rg_{cd}^{(n)}h^{cd}\right)+\nn\\
&&+ \frac{1}{G^{d/2}}\partial_r \left( G^{d/2}\partial_r g_{cd}^{(n)}u^cu^d\right)+\nn\\
&&+\frac{d}{2G^{d/2+1}}\partial_r\left( G^{d/2}G^{\prime}g_{cd}^{(n)}u^cu^d\right)+\nn\\
&&-\frac{d}{2}\frac{G^{\prime 2}}{G^2}g_{cd}^{(n)}u^cu^d, 
\ea
where we defined
\be
\delta R^{(n)} = g^{(0) ab} \delta R^{(n)}_{ab}. 
\ee
Notice that 
\ba\label{dirichlet}
&&\lambda^{-2}\delta R_{a\mu}^{(n)}n^{\mu}|_{\Sigma_c}=\\
&&\quad u_au^cu^dg^{(n)}_{cd}|_{\Sigma_c}\left( \frac{d}{4}F'(r_c)G'(r_c)+F''(r_c)   \right)+\nn\\
&&
 -h^{c}_au^dg^{(n)}_{cd}|_{\Sigma_{c}}\!\!\left( \!G''(r_c)\!+\!\frac{(d-2)}{4}G'(r_c)^2\!+\!\frac{1}{2}F'(r_c)G'(r_c)\!      \right),\nn
\ea
which is identically zero when we impose the required Dirichlet boundary conditions on the hypersurface $\Sigma_c$, i.e. $g_{ab}^{(n)}|_{\Sigma_c}=0$ with $n\geq 1$.

The Einstein equations can be used to solve for the different components of the metric perturbations,
which at each gradient order can be decomposed in terms of a basis of linearly independent scalars, vectors and traceless symmetric tensors
\be\label{eq:expansionmetric}
g_{ab}^{(n)}=\alpha^{(n)}u_au_b+2u_{(a}\beta_{b)}^{(n)}+\tilde{\gamma}_{ab}^{(n)}+\frac{1}{d}\gamma ^{(n)}h_{ab}
\ee
with 
\be
u^a\beta_{a}^{(n)}=u^a \tilde{\gamma}_{ab}^{(n)}=\tilde{\gamma}^{(n)}_{ab}h^{ab}=0.
\ee

Using  (\ref{integrability})  it is easy to show that the following equations must be satisfied 
\ba
&& \frac{1}{2}\partial_r^2\left(\frac{1}{G}\gamma^{(n)}\right)+\frac{1}{2}\frac{G^\prime}{G}\partial_r\left(\frac{1}{G}\gamma^{(n)}\right)+\bar{\TT}^{(n)}_{rr}=0,\label{tracegamma}\\
&& \frac{\lambda}{2}\partial_r^2\beta_a^{(n)}\!+\!\frac{\lambda}{4}(d\!-\!2)\frac{G^\prime}{G}\partial_r\beta_a^{(n)}\!-\!\lambda\left( \frac{1}{4}(d\!-\!2)\frac{G^{\prime 2}}{G^2}\!+\!\frac{1}{2}\frac{G^{\prime\prime}}{G}\right)\beta_a^{(n)}\!+\nn\\
&&
\quad-(\!\hat{R}^{(n)}_{rb} -\bar{\TT}^{(n)}_{rb}) h^b_a=0,\label{beta}\\
&&\frac{d}{2}\lambda^2\frac{1}{G^{d/2}}\partial_r\left(G^{d/2}G^\prime\alpha^{(n)}\right)-\frac{d}{2}\lambda^2\frac{G^{\prime 2}}{G}\alpha^{(n)}+\nn\\
&&\quad-\frac{\lambda^2}{2G^{d/2}}\partial_r\left(G^{d/2}F\partial_r\gamma^{(n)}\right)+\lambda^2\frac{G^\prime F}{G}\partial_r\gamma^{(n)}-\frac{\lambda^2}{2}\frac{FG^{\prime 2}}{G^2}\gamma^{(n)}+\nn\\
&&
\quad-\frac{d}{4}\lambda^2FG^\prime\partial_r\left(\frac{1}{G}\gamma^{(n)}\right)+\left(\hat{R}_{ab}^{(n)}-\bar{\TT}_{ab}^{(n)}\right)h^{ab}=0,\label{alpha} \\
&&
\frac{\lambda^2}{2}F\partial_r^2\tilde{\gamma}_{ab}^{(n)}+\lambda^2\left( \frac{1}{4}(d-4)\frac{FG^\prime}{G}+\frac{1}{2}F^\prime\right)\partial_r\tilde{\gamma}_{ab}^{(n)}+\nn\\
&&
\quad +\frac{\lambda^2}{2}\frac{FG^{\prime 2}}{G^2}\tilde{\gamma}_{ab}^{(n)}-\hat{R}_{cd}^{(n)}h^c_ah^d_b+\frac{1}{d}\hat{R}_{cd}^{(n)}h^{cd}h_{ab}+ \\
&&\quad +\left(\bar{\TT}_{cd}^{(n)}h^c_ah^d_b-\frac{1}{d}\bar{\TT}_{cd}^{(n)}h^{cd}h_{ab}\right)=0.\label{gamma} \nn
\ea
These equations are, respectively, the $(rr)$ equation; the $h^{ab}$ projection of the $(ra)$ equation; the $h^{ab}$ trace of the $(ab)$ equation and the projection $(h^{c}_{a} h^{d}_{b} -1/d\, h^{cd} h_{ab})$ of the $(ab)$ equation. Additional, linearly dependent, equations may be obtained from other projections of the Einstein equations. 
For example, projecting the $(ra)$ components of the Einstein equations onto $u^a$ one obtains an additional equation for $\alpha^{(n)}$
\ba
&&\frac{1}{2} \lambda \partial_r^2 \alpha^{(n)} + \lambda \frac{d G^\prime}{4 G} \partial_r \alpha^{(n)} - 
\frac{\lambda}{4} F^{\prime} \partial_r \left ( \frac{\gamma^{(n)}}{G} \right ) +\nn\\
&&
\qquad + (\hat{R}^{(n)}_{ra} - \bar{\cal T}^{(n)}_{ra}) 
u^a = 0. 
\ea
Since this equation is second order it is more convenient to use the first order equation \eqref{alpha}.
Similarly one can obtain an additional equation for $\beta^{(n)}$ from the projection $h_{b}^{c} u^a$ of the 
$(ab)$ components of the Einstein equations. 

The Bianchi identities  at order $n$  are
\be
(\nabla_{\mu}G^{\mu\nu})^{(n)}=\sum_{k=0}^{n}\nabla_{\mu}^{(n-k)}G^{(k)\mu\nu }=0.
\ee
Using the Einstein equations at each order and the conservation of the bulk stress energy tensor at order $n$,  $(\nabla_{\mu}\TT^{\mu\nu})^{(n)}=0$,  the following identity holds
\be
\nabla_{\mu}^{(0)}G^{(n)\mu\nu}=\nabla_{\mu}^{(0)}\TT ^{(n)\mu\nu},
\ee
or equivalently
\be\label{generalbianchi}
\nabla_{\mu}^{(0)}R^{(n)\mu\nu}=\nabla_{\mu}^{(0)}\bar{\TT}^{(n)\mu\nu}.
\ee
The independent $(r)$ and $(a)$ components of (\ref{generalbianchi}) are 
\ba 
&&\frac{1}{2}\lambda ^4FF'(R_{rr}^{(n)}-\bar{\TT}_{rr}^{(n)})+\lambda ^3F'(R_{ra}^{(n)}-\bar{\TT}_{ra}^{(n)})u^a+\label{bianchia}\\
&&
-\frac{1}{2}\lambda ^2\frac{FG'}{G^2}(R_{ab}^{(n)}-\bar{\TT}_{ab}^{(n)})h^{ab}+\nn\\
&&+\frac{\lambda ^2F}{G^{d/2}}\partial_r\left(G^{d/2}(\lambda ^2F(R_{rr}^{(n)}\!-\!\bar{\TT}_{rr}^{(n)})\!+\!2\lambda (R_{ra}^{(n)}\!-\!\bar{\TT}_{ra}^{(n)})u^a)\right)\! +\nn\\
&&+\lambda^2\frac{1}{G^{d/2}}\partial_r\left( G^{d/2}(R_{ab}^{(n)}-\bar{\TT}_{ab}^{(n)})u^au^b\right)=0,\nn \\
&&\frac{1}{G^{d/2}}\partial_r\left(G^{d/2}\left( \lambda F(R_{ar}^{(n)}-\bar{\TT}_{ar}^{(n)})+(R_{ab}^{(n)}-\bar{\TT}_{ab}^{(n)})u^b\right) \right)=0\nn.
\ea
From (\ref{generaleq}) and  using the leading order equations of motion (\ref{eqmotion0}) one can show that the following expressions are identically satisfied
\ba
 &&\lambda F\,\delta R_{ar}^{(n)}u^a+\delta R_{ab}^{(n)}u^au^b=\nn\\
 &&\qquad\qquad\left(-\frac{1}{d}\TT^{(0)}+\lambda\TT^{(0)}_{re}u^e\right)g_{ab}^{(n)}u^au^b,\nn\\
  &&\lambda F\,\delta R_{ar}^{(n)}h^a_c+\delta R_{ab}^{(n)}u^bh^a_c=\nn\\
  &&\qquad\qquad
  \left( -\frac{1}{d}\TT^{(0)}+\frac{1}{d}\frac{1}{G}\TT_{ef}^{(0)}h^{ef} \right)g_{ab}^{(n)}u^bh^a_c. 
\ea
Using
\ba
&&\lambda F\,\bar{\TT}_{ar}^{(n)}+\bar{\TT}_{ab}^{(n)}u^b=\lambda F\, \TT_{ar}^{(n)}+\TT_{ab}^{(n)}u^b+\\
&&
\qquad \qquad \qquad -\frac{1}{d}\TT^{(0)}g_{ab}^{(n)}u^b-\frac{1}{d}\sum_{k=1}^{n-1}g_{ab}^{(k)}u^b\TT^{(n-k)}\nn
\ea
equations (\ref{bianchia}) can be brought into the form:
\ba\label{integrability2}
&&
\partial_r\Big(G^{d/2}( \lambda F(\hat{R}_{ar}^{(n)}\!\!-\!\! \TT_{ar}^{(n)})+(\hat{R}_{ab}^{(n)}\!\!-\!\!\TT_{ab}^{(n)})u^b +\nn\\
&&
\qquad +\lambda\TT^{(0)}_{re}u^eg_{ab}^{(n)}u^b+\frac{1}{d}\sum_{k=1}^{n-1}g_{ab}^{(k)}u^b\TT^{(n-k)}) \Big)u^a=0,\nn\\
&&
\partial_r\Big(G^{d/2}( \lambda F(\hat{R}_{ar}^{(n)}\!\!-\!\! \TT_{ar}^{(n)})+(\hat{R}_{ab}^{(n)}\!\!-\!\!\TT_{ab}^{(n)})u^b +\nn\\
&&
\qquad +\frac{1}{dG}\TT^{(0)}_{ef}h^{ef}g_{ab}^{(n)}u^b+\frac{1}{d}\sum_{k=1}^{n-1}g_{ab}^{(k)}u^b\TT^{(n-k)}) \Big)h^a_c=0.\nn\\
&&
\ea
Integrating these conditions and evaluating them on $\Sigma_c$ gives
\be
 \left(\lambda (\hat{R}_{ar}^{(n)}\!\!-\!\! \TT_{ar}^{(n)})+(\hat{R}_{ab}^{(n)}\!\!-\!\!\TT_{ab}^{(n)})u^b\right)|_{\Sigma_c}={f}_a^{(n)}(x)|_{\Sigma_c},
\ee
where ${f}^{(n)}_{a} (x)$ arises as an integration constant.

The Gauss-Codazzi equations on $\Sigma_c$ at order $n$ are given by
\ba\label{consBY} 
\nabla^{b}T_{ab}^{(n){\rm BY}}|_{\Sigma_c}&=&-2R_{a\mu}^{(n)}n^{\mu}|_{\Sigma_{c}}=-2(\hat{R}_{a\mu}^{(n)}+\delta R_{a\mu}^{(n)})|_{\Sigma_c}n^{\mu} =\nn\\
&=&-2\left(\lambda\hat{R}^{(n)}_{ar}+\hat{R}_{ab}^{(n)}u^b \right)|_{\Sigma_c}= \nonumber \\ 
&=& - 2 n^{\mu} {\cal T}^{(n)}_{a \mu} -2 f_a^{(n)}(x)|_{\Sigma_c}. 
\ea
As discussed around \eqref{bulkconstraint}, conservation of the fluid stress tensor requires that ${\cal T}_{a \mu} n^{\mu}$ vanishes to all orders, which in turn requires that ${\cal T}^{(n)}_{a \mu}n^{\mu} = 0$ to all orders $n$ since $n^{\mu}$ does not change due to the required Dirichlet boundary conditions. If the fluid is not conserved then  ${\cal T}^{(n)}_{a \mu}n^{\mu} \neq 0$ characterises this non-conservation. In both cases the integration constant arising from integrating the Bianchi identities is therefore zero for the fluid stress energy tensor to satisfy the required conservation equation.

\bigskip
The extrinsic curvature at order $n$ is given by 
\be 
K_{ab}^{(n)}=\frac{1}{2}(\LL_{n}g_{ab})|_{\Sigma_c}^{(n)}=\hat{K}^{(n)}_{ab}+\delta K^{(n)}_{ab},
\ee
where again we divide it into two contributions: $\hat{K}_{ab}^{(n)}$ corresponding to contributions at order $n$ coming from spacetime derivatives $\partial_a$ of $g^{(n-1)}$ and $\delta K^{(n)}_{ab}$ coming from radial derivatives of $ g^{(n)}$. Let us introduce 
the following notation for the velocity derivatives 
\ba
\theta=\partial_c u^c;&\qquad& a_a=Du_a;\\ 
\KK_{ab}=h_{(a}^ch_{b)}^d\partial_{c}u_{d};&\qquad&
 \sigma_{ab}=\KK_{ab}-\frac{1}{d}\theta h_{ab},
\ea
where as defined earlier, $D = u^{c} \nabla_{c}$.
We obtain 
\ba\label{eq:hatK}
\hat{K}^{(1)}_{ab} &=&\sigma_{ab}+\frac{1}{d}\theta h_{ab}-u_{(a}a_{b)}-u_{(a}\partial_{b)}\ln \lambda,\\
\hat{K}^{(n)}_{ab}&=&\frac{1}{2}Dg_{ab}^{(n-1)}|_{\Sigma_c}\qquad\text{with}\qquad n>1, 
\ea
where  we have used the Dirichlet boundary condition
and  the expression for the normal vector
(\ref{normalvector}), and
\ba
\delta K_{ab}^{(n)}&=&\frac{1}{2}\lambda\,\partial_rg_{ab}^{(n)}|_{\Sigma_c}=\frac{1}{2}\lambda\Big( \alpha^{(n)'}(r_c)u_au_b+\\
&&+2\beta_{(a}^{(n)'}(r_c)u_{b)}+\tilde{\gamma}_{ab}^{(n)'}(r_c)+\frac{1}{d}\gamma^{(n)'}(r_c)h_{ab}\Big)\nn
\ea
The Brown-York stress energy tensor can also be split as
\be\label{eq:BY}
T_{ab}^{{\rm BY}(n)}=\hat{T}^{{\rm BY}(n)}_{ab}+\delta T_{ab}^{{\rm BY}(n)}.
\ee
where again 
\be\label{eq:BYn-1}
\hat{T}_{ab}^{{\rm BY}(n)}=2(\hat{K}^{(n)}\eta_{ab}-\hat{K}_{ab}^{(n)}),
\ee
with
\be
\hat{K}^{(n)}=\hat{K}_{ab}^{(n)}\eta^{ab},
\ee
is the contribution coming from the $(n-1)$th order metric, and 
\be\label{eq:BYn}
\delta T_{ab}^{{\rm BY}(n)}=2(\delta K_{ab}^{(n)}\eta_{ab}-\delta K_{ab}^{(n)}),
\ee
with
\be
\delta K^{(n)}=\delta K_{ab}^{(n)}\eta^{ab}, 
\ee
is the contribution coming from the correction of the  metric at order $n$.
The last part can be worked out formally giving
\ba
&&\lambda^{-1}\delta T_{ab}^{{\rm BY}(n)}=\Big(-u_a u_b\gamma^{(n)\prime}(r_c)-2\beta_{(a}^{(n)\prime}(r_c)u_{b)}+\nn\\
&&
+h_{ab}\big(\!-\!\alpha^{(n)\prime}(r_c)\!+\!\frac{(d-1)}{d}\gamma^{(n)\prime}(r_c)\big)
-\tilde{\gamma}_{ab}^{(n)\prime}(r_c)\Big)
\ea
We will work in Landau gauge for the fluid stress tensor 
\be\label{Landauformal}
T_{ab}^{{\rm F} (n)} u^a = 0;\quad n\ge 1.
\ee
Imposing this condition on the fluid stress energy tensor given in (\ref{fluidstress}), gives two additional constraints  order by order $n\ge 1$ on the Brown-York stress energy tensor
\ba\label{landau}
\lambda\gamma^{(n)\prime}(r_c)-\hat{T}_{ab}^{{\rm BY}(n)}u^au^b&=&0, \\
\lambda\beta_c^{(n)\prime}(r_c)+\hat{T}_{ab}^{{\rm BY}(n)}u^ah^b_c&=&0, \nn
\ea
which follow from the two independent projections of the condition above (\ref{Landauformal}), and we used the fact that $C_1(r_c)=1$ for a flat metric on $\Sigma_c$ and that $C_2(r_c)$ results only in a redefinition of the pressure and energy density.  

\subsection{General first order hydrodynamics}

The Ricci tensor contribution coming from the seed metric  is
\ba\label{Riccifirst}
\hat{R}^{(1)}_{rr}&=&0, \\
\hat{R}^{(1)}_{ra}&=&u_a\left( \frac{G'}{2G}\partial_cu^c+\frac{d}{2G}DG'-\frac{dG'}{4G^2}DG\right)+\nn\\
&&
+\frac{d}{4}\frac{G'}{G}a_a-(d-1)\frac{1}{2G}D_a^{\perp}G'+\nn\\
&&
+(d-1)\frac{G'}{2G^2}D_a^{\perp}G-\frac{1}{4}d\frac{G^\prime}{G}D_a^{\perp}\ln \lambda,\nonumber\\ 
\lambda^{-1}\hat{R}^{(1)}_{ab}&=&u_au_b\Big( \frac{1}{2}F'\partial_cu^c-\frac{dG'}{4G}DF+\nn\\
&&\quad+\frac{dF'}{4G}DG-\frac{1}{2}d\frac{G^\prime F}{G}D\ln \lambda\Big)+\nonumber\\
&&+h_{ab}\left( -\frac{1}{2}G'\partial_cu^c -DG'-\frac{(d-2)}{2}\frac{G'}{G}DG    \right)+\nonumber\\
&&+u_{(a}a_{b)}\left( F'+\frac{(d-2)}{2}\frac{G'F}{G}      \right)+\nonumber\\
&&+u_{(a}D_{b)}^{\perp}F'+\frac{(d-2)}{2}\frac{G'}{G}u_{(a}D^{\perp}_{b)}F-\frac{d}{2}G'\KK _{ab}+\nonumber\\
&&+u_{(a}D^{\perp}_{b)}\ln \lambda\left( F^\prime +\frac{1}{2}(d-2)\frac{G^\prime F}{G}\right) . \nn
\ea
Recalling that 
\be
G(r_c)=F(r_c)=1;\qquad  DG(r_c)=DF(r_c)=0, 
\ee
the momentum  constraint $R_{a\mu}^{(1)}n^{\mu}|_{\Sigma_c}=(\hat{R}_{a\mu}^{(1)}+\delta R_{a\mu}^{(1)})n^{\mu}|_{\Sigma_c}=0$ evaluates into the ideal fluid equations of motion (\ref{cons}).

The Brown-York stress energy tensor contribution (\ref{eq:BYn-1}) arising  from the seed metric (\ref{form1a}) is given by
\ba
\hat{T}_{ab}^{{\rm BY}(1)}&=&-2\theta u_au_b
+2h_{ab}\left( \frac{(d-1)}{d}\,\theta-D\ln\lambda \right)+\nn\\
&&-2\sigma_{ab}+2u_{(a}a_{b)}+2u_{(a}D^{\perp}_{b)}\ln\lambda,
\ea
and the Landau gauge condition (\ref{landau}) becomes
\ba\label{landau2}
\lambda\gamma^{(1)\prime}(r_c)&=&-2\theta,\\
\lambda\beta_a^{(1)\prime}(r_c)&=&a_a+D_a^{\perp}\ln\lambda. \nn
\ea
Hence the complete contribution to the Brown-York stress energy tensor (\ref{eq:BY}) at first order in Landau gauge is given by
\be\label{BYfirst}
T_{ab}^{{\rm BY}(1)}=-h_{ab}\left(2\,D\ln\lambda +\lambda\alpha^{(1)\prime}(r_c)\right)-2\sigma_{ab}-\lambda\tilde{\gamma}_{ab}^{(1)\prime}(r_c).
\ee
Note that this result holds generally, regardless of the structure of the bulk stress energy tensor.

Consider first the case in which there is no matter. Then to first order in gradients there is only one independent  scalar which may be chosen to be $\theta$; there is only one vector orthogonal to $u^a$, which we may choose as $a_a$; the only symmetric traceless tensor orthogonal to $u^a$ is $\sigma_{ab}$. The other scalars ($D \rho$, $D \ln p$) can be eliminated using the equation of state and the fluid equations, while the other vector $D^{\perp}_a \ln p$ can similarly be eliminated. 
Hence 
the dissipative part of the fluid stress energy tensor can be written as
\be\label{Fluidfirst}
T_{ab}^{\rm F (1)} = - 2 \eta(r_c) \sigma_{ab} - \xi(r_c) \theta h_{ab},
\ee
to first order in gradients, where
\ba
\eta(r_c)\sigma_{ab}&=&\left(\frac{1}{2}\lambda h_{ac} h_{bd} \tilde{\gamma}^{cd (1)\prime}(r_c)+\sigma_{ab} \right),\\
\xi(r_c)\theta &=&\lambda\alpha^{(1)\prime}(r_c)+2D\ln\lambda,
\ea
define the shear viscosity $\eta(r_c)$ and the bulk viscosity $\xi(r_c)$ at the hypersurface $\Sigma_c$.

Notice that the shear over entropy ratio
\be
\frac{\eta(r_c)}{s(r_c)}\sigma_{ab}=\frac{\frac{1}{2}\lambda h_{ac} h_{bd} \tilde{\gamma}^{cd (1)\prime}(r_c)+\sigma_{ab} }{4\pi G(r_H)^{d/2}},
\ee
where we used (\ref{entr}), satisfies the universal value $\eta/s=1/4\pi$ only if
\be
\frac{1}{2}\lambda h_{ac} h_{bd} \tilde{\gamma}^{cd (1)\prime}(r_c)+\sigma_{ab}=G(r_H)^{d/2} \sigma_{ab}.
\ee

\subsection{The cutoff AdS fluid, revisited}

We will now consider the case of a negative cosmological constant, expanding around the asymptotically $AdS$ black brane geometry (\ref{bb}). The first step is to write the latter in the required form (\ref{form1a}) through (\ref{GFbb}).
One computes the hydrodynamic solution by promoting $r_{H}$ and $u_a$ to be slowly varying $r_H(x)$, $u_a(x)$. Since $G$ is independent of both quantities $DG = D^{\perp} G = 0$, which considerably simplifies the formulae given above. 

The bulk stress energy tensor is
\be
\TT_{\mu\nu}=-\Lambda g_{\mu\nu}.
\ee
The condition for the Brown-York stress energy tensor (and hence the fluid stress energy tensor) to be conserved is identically satisfied at zeroth order since $\TT_{\mu\nu}^{(0)}n^{\mu}=0$. Conservation at higher orders is guaranteed by the Dirichlet condition for the metric on $\Sigma_c$.

The reduced bulk stress energy tensor is
\be
\bar{\TT}^{(1)}_{\mu\nu} =\frac{2}{d}\Lambda g_{\mu\nu}^{(1)}
\ee
 which leads to
\ba
&& \bar{\TT}_{rr}^{(1)}=0;\qquad
\bar{\TT}_{ab}^{(1)}h^{ab}=\frac{2}{d}\Lambda\gamma^{(1)};\\
&& \bar{\TT}_{ra}^{(1)}h^a_b=0;\qquad
\bar{\TT}_{ab}^{(1)}h^a_ch^b_d-\frac{1}{d}\bar{\TT}_{ab}^{(1)}h^{ab}h_{cd}=\frac{2}{d}\Lambda\tilde{\gamma}^{(1)}_{cd}. \nn
\ea
Using the conservation of the stress energy tensor (\ref{cons}) it is possible to trade  derivatives of the horizon radius $r_H$  with  derivatives of the fluid velocities 
\be\label{derhor}
\frac{\partial_a r_H}{r_H}=\left(\frac{1}{d}\theta u_a-\delta(r_c)a_a\right),
\ee
where
\be
\delta(r_c)=\frac{2(1-(r_H/r_c)^{d+1})}{2+(d-1)(r_H/r_c)^{d+1}}=\frac{2\,h_c}{2h_c+(d+1)(1-h_c)},
\ee
notice that when $r_c\rightarrow\infty$ we have $\delta(r_c)\rightarrow 1$ and one recovers the usual relation for conformal fluids (\ref{CFTeom}).
Equations (\ref{tracegamma}-\ref{gamma}) are
\ba
&&\partial_r^2\left(\frac{r_c^2}{r^2}\gamma^{(1)}\right)+\frac{2}{r}\partial_r\left(\frac{r_c^2}{r^2}\gamma^{(1)}\right)=0,\label{gammatracefirst}\\
&&d\,h_c\frac{r}{L^2}\alpha^{(1)\prime}+h_c\frac{d(d-1)}{L^2}\alpha^{(1)}+\nn\\
&&\,\,+\!\left( \frac{(d+1)}{L^2}\!+\!\frac{(d-2)}{L^2}h\right)\gamma^{(1)}
\!-\!\frac{r}{L^2}\left((d-1)h\!+\!\frac{1}{2}rh^\prime \right)\gamma^{(1)\prime}\!+\!\nn\\
&&\,\,-\frac{1}{2}\frac{r^2}{L^2}h\gamma^{(1)\prime\prime}-2d\frac{r}{r_cL}\sqrt{h_c}\theta
 =0,\label{alphafirst}\\
&&\beta_a^{(1)\prime\prime}+\frac{(d-2)}{r}\beta_a^{(1)\prime}-2\frac{(d-1)}{r^2}\beta_a^{(1)} +\nn\\
&&\quad-\frac{d}{r}\frac{L}{r_c\sqrt{h_c}}\delta(r_c)a_a=0,\label{betafirst}\\
&&\partial_r\left( r^{d+2} h(r)\partial_r\left(\tilde{\gamma}^{(1)}_{ab}\frac{L^2}{r^2}\right)\right)+2 d \sqrt{h_c}\frac{L^3 r^{d-1}}{r_c}\sigma_{ab}=0.\label{gammafirst}
\ea
The Landau gauge condition (\ref{landau2}) on the stress tensor gives the following  equations
\ba
&&\gamma^{(1)\prime}(r_c)+\frac{2L}{r_c\sqrt{h_c}}\theta =0,\label{landaufirst1}\\
&&\beta_a^{(1)\prime}(r_c)\!=\!\frac{2L}{r_c\sqrt{h_c}}\left(\frac{h_c+(d+1)(1-h_c)}{2h_c+(d+1)(1-h_c)}\right)a_a\label{landaufirst2}.
\ea
Solutions to equations (\ref{gammatracefirst}-\ref{gammafirst}) with Dirichlet boundary conditions $\gamma^{(n)}(r_c)=\alpha^{(n)}(r_c)=\beta_a^{(n)}(r_c)=\tilde{\gamma}_{ab}^{(n)}(r_c)=0$ are given by
\ba
\gamma^{(1)}&=&\frac{2L}{r_c\sqrt{h_c}}r\left( 1-\frac{r}{r_c} \right)\theta ,\label{tracegammasol}\\
\alpha^{(1)}&=&\frac{L}{d\,h_c^{3/2}}\frac{r}{r_c}\Big(-(d+1)\left(1-\frac{r_c^d}{r^d}\right)+(d-1)h+\nn\\
&&\qquad +\left(2-(d+1)\frac{r_c^d}{r^d}\right)h_c\Big) \theta,\\
\beta^{(1)}_a&=&-\frac{L\delta(r_c)}{r_c^2h_c^{3/2}}r\left(r \left(\frac{r_c^{d+1}}{r^{d+1}}-1\right)+r_c\left( 1-\frac{r_c^d}{r^d}\right)h_c\right)a_a,\nn\\
&&\\
\tilde{\gamma}_{ab}^{(1)}&=&\frac{2L}{r_cr_H}\sqrt{h_c}r^2\left(k(r)-k(r_c)+\frac{1}{(d+1)}\log h/h_c\right)\sigma_{ab},\nn\\
&&\label{gammasol}
\ea
where
\be\label{hyper}
k(r)=\frac{r_H}{r}\, _{2}F_1\left(1,\frac{1}{d+1},1+\frac{1}{d+1},\frac{r_H^{d+1}}{r^{d+1}}\right),
\ee
see appendix \ref{appendix1} for details of the derivation.
The fluid stress energy tensor (\ref{Fluidfirst}) at first order  is
\be
T_{ab}^{{\rm F}(1)}= -2\eta(r_c)\sigma_{ab}-\xi(r_c)\theta h_{ab},
\ee
with  shear and bulk viscosity  given by 
\ba
&&\eta(r_c)\sigma_{ab}\!=\!\left(\frac{1}{2}\frac{r_c}{L}\sqrt{h_c}\gamma^{(1)\prime}_{ab}(r_c)+\sigma_{ab}\right)\!=\!\frac{r_H^d}{r_c^d}\sigma_{ab},\label{shear1}\\
&&\xi(r_c)\!=\! \frac{(d+1)}{d}\frac{(1-h_c)}{h_c}+\frac{r_c}{L}\sqrt{h_c}\alpha^{(1)\prime}(r_c)=0.\label{bulk1}
\ea
Hence, although the fluid is non conformal due to non conformal equation of state (\ref{eqstate}) giving a non zero trace for the stress energy tensor, the bulk viscosity is vanishing.
Given the entropy density $s(r_c)=4\pi r_H^d/r_c^d$ the universal shear over entropy ratio bound is recovered at each hypersurface $\Sigma_c$
\be 
\frac{\eta(r_c)}{s(r_c)}=\frac{1}{4\pi},
\ee
confirming the results found previously  for the non relativistic fluid dual to a finite cutoff hypersurface in AdS gravity \cite{Cai:2011xv} and the relativistic version of it \cite{Brattan:2011my}, see also \cite{Emparan:2013ila}. Alternative derivations using RG flows can be found in \cite{Bredberg:2010ky,Kuperstein:2011fn,Kuperstein:2013hqa},
see also \cite{Iqbal:2008by} for a derivation using linear response theory.

\subsection{Relation to the conformal fluid}

The finite cutoff solution can be connected to the usual AdS/CFT results if we first redefine the field theory coordinates
\be
y^a=\frac{L}{r_c}x^a
\ee 
in order to have a conformally flat metric on the boundary.
The leading order metric now reads
\be\label{blackbranemetric2}
ds^2=-2\frac{u_a}{\sqrt{h_c}}dy^adr+\frac{r^2}{L^2}\left(h_{ab}-h/h_cu_au_b\right)dy^ady^b,
\ee
so that as $r_c\rightarrow\infty$ we have $h_c\rightarrow 1$ and the metric is the usual black-brane metric.
The rescaling acts on the derivatives $\partial_a^x=\frac{L}{r_c}\partial_a ^y$ with $u_a$ unchanged, hence we have
\be
\theta ^x=\frac{L}{r_c}\theta ^y;\quad a_a^x=\frac{L}{r_c}a_a^y;\quad \sigma_{ab}^x=\frac{L}{r_c}\sigma_{ab}^y, 
\ee
and the perturbations (\ref{tracegammasol}-\ref{gammasol}) in the rescaled coordinates where we take  into account $dx^adx^b\rightarrow r_c^2/L^2 dy^ady^b$ now read
\ba
\gamma^{y(1)}&=&\frac{2}{\sqrt{h_c}}r\left( 1-\frac{r}{r_c} \right)\theta^y ,\label{tracegammasol2}\\
\alpha^{y(1)}&=&\frac{1}{d\,h_c^{3/2}}r\Big(-(d+1)\left(1-\frac{r_c^d}{r^d}\right)+(d-1)h+\nn\\
&&+\left(2-(d+1)\frac{r_c^d}{r^d}\right)h_c\Big) \theta^y,\\
\beta^{y(1)}_a&=&-\frac{\delta(r_c)}{r_ch_c^{3/2}}r\left(r \left(\frac{r_c^{d+1}}{r^{d+1}}-1\right)+r_c\left( 1-\frac{r_c^d}{r^d}\right)h_c\right)a_a^y,\nn\\
&&\\
\tilde{\gamma}_{ab}^{y(1)}&=&\frac{2}{r_H}\sqrt{h_c}r^2\left(k(r)-k(r_c)+\frac{1}{(d+1)}\log h/h_c\right)\sigma_{ab}^y.\nn\\
&&\label{gammasol2}
\ea
The limit $r_c\rightarrow\infty$ gives now
\ba
\gamma^{y(1)}&\rightarrow& 2r\theta^y ,\label{pert1}\\
\alpha^{y(1)}&\rightarrow&\frac{r}{d}(d-1)(h-1) \theta ^y\\
\beta^{y(1)}_a&\rightarrow& -ra_a^y,\\
\tilde{\gamma}_{ab}^{y(1)}&\rightarrow&2\frac{r^2}{r_H}\left(k(r)-k(\infty)+\frac{1}{(d+1)}\log h\right)\sigma_{ab}^y=\nn\\
&&=2\frac{r^2}{r_H}F(r/r_H)\sigma_{ab}^y\label{pert4}.
\ea
where asymptotically $F(x)\sim \frac{1}{x}-\frac{1}{(d+1)}\frac{1}{x^{d+1}}$.
Notice that the first order perturbation of the AdS black brane metric is given in 
\cite{Bhattacharyya:2008jc} as
\ba\label{Min}
ds^{(1)2}&=&2\frac{r^2}{r_H}F(r/r_H)\sigma_{ab}dy^ady^b+\frac{2}{d}r\theta^y u_au_bdy^ady^b+\nn\\
&&-2ra^y_{(a}u_{b)}dy^ady^b.
\ea
Relative to the metric perturbation given above the coefficients in the vector and tensor sector agree but those in the scalar sector do not. However, the two metrics are connected by the following diffeomorphism: consider a shift of the radial coordinate $r$ such that
\be
r  \rightarrow r - \frac{1}{d} \theta^y.
\ee
Working to first order in gradients this results in a shift
\be\label{redefinition}
\delta \gamma^{y (1)} = - 2 r \theta^y; \qquad
\delta \alpha^{y (1)} = \frac{r \theta^y}{d} \left (2 - (d-1)(h-1) \right ), 
\ee
thereby bringing the metric into the form given in \cite{Bhattacharyya:2008jc}.

The shear and bulk viscosity of the dual theory on the boundary can be obtained directly from  (\ref{shear1}-\ref{bulk1}) according to the prescription (\ref{prescription}) by\footnote{where we take $L=1$ for notational simplicity.}
\ba
\eta^{\rm CFT}\sigma_{ab}^y&=&\lim_{r_c\to \infty}r_c^{d-1}\eta(r_c)r_c\sigma_{ab}^y=r_H^d\sigma_{ab}^y,\\
\xi^{\rm CFT}&=& \lim_{r_c\to\infty}r_c^{d-1}\xi(r_c)=0. \nn
\ea
Hence, the stress energy tensor at finite cutoff reproduces the usual AdS/CFT results as the cutoff is taken to the boundary.

The same result can be obtained  directly from the metric perturbation at infinity (\ref{pert1}-\ref{pert4}) and using the  usual AdS/CFT prescription (\ref{adscftprescription}). 
Since the metric on the boundary is (conformally) flat the dual fluid stress-energy tensor is
\be
\langle T_{ab}^{\rm CFT}\rangle =\lim_{R\to\infty}R^{d-1}(T_{ab}^{\rm BY}-2d \gamma_{ab}),
\ee
where the Brown-York stress-energy tensor is given by
\ba
T_{ab}^{{\rm BY}(0)}&=&2\left(K^{(0)}\gamma_{ab}^{(0)}-K_{ab}^{(0)}\right),\\
T_{ab}^{{\rm BY}(1)}&=&2\left( K^{(0)}\gamma_{ab}^{(1)}+K^{(1)}\gamma_{ab}^{(0)}-K_{ab}^{(1)}\right), \nn
\ea
and
 the extrinsic curvature is computed at a finite cutoff surface $\Sigma_R$ giving
\ba
K_{ab}^{(0)}&=&\frac{1}{2}n^{r(0)}\partial_r g_{ab}^{(0)}|_R,\\
K_{ab}^{(1)}&=&\frac{1}{2}n^{r(0)}\partial_r g_{ab}^{(1)}|_R+\frac{1}{2}n^{r(1)}\partial_r g_{ab}^{(0)}|_R+\nn\\
&&+\frac{1}{2}n^{c(0)}\partial_c g_{ab}^{(0)}|_R+g_{(a\mu}\partial_{b)}^{(0)}n^{\mu(0)}|_R,\nn \\
K^{(0)}&=&K_{ab}^{(0)}\gamma^{ab(0)},\nn \\ K^{(1)}&=&K_{ab}^{(1)}\gamma^{ab(0)}+K_{ab}^{(0)}\gamma^{ab(1)}. \nn
\ea
The induced metric on the hypersurface $\Sigma_R$  is generically non flat since
\ba
\gamma_{ab}^{(0)}&=&R^2\left(h_{ab}-h(R)u_au_b\right),\\
\gamma_{ab}^{(1)}&=&\frac{(d-1)}{d}\left(h(R)-1\right)R\, \theta ^yu_au_b-2R\,a_{(a}^yu_{b)}+\nn\\
&&+2\frac{R^2}{r_H}F(R/r_H)\sigma_{ab}^y+ \frac{2}{d}R\,\theta^y h_{ab},\nn \\
\gamma^{ab(1)}&=&-\gamma^{ac(0)}\gamma^{bd(0)}\gamma_{cd}^{(1)}, \nn
\ea
giving non trivial contributions to the normal vector up to and including order one
\ba
&&\vec{n}^{(0)}=R\sqrt{h(R)}\partial_r +\frac{1}{R\sqrt{h(R)}}u^a\partial_a,\\
&&\vec{n}^{(1)}=-\frac{1}{2}\frac{(d-1)}{d}\frac{h(R)-1}{\sqrt{h(R)}}\theta^{y}\partial_r+\nn\\
&&\quad +\frac{1}{R^2\sqrt{h(R)}}\left(\frac{1}{2}\frac{(d-1)}{d}\frac{(h(R)-1)}{h(R)}\theta^{y} u^a-a^{(y)a}\right)\partial_a.\nn
\ea
Considering everything together and using the leading order equations of motion to relate derivatives of the horizon radius $r_H$ to derivatives of $u_a$ 
\be\label{CFTeom}
\frac{\partial_{a}r_H}{r_H}=\frac{1}{d}\theta^y u_a-a_a,
\ee
we obtain the fluid stress-energy tensor dual to a black-brane metric in  $AdS$ spacetime 
\be
\langle T_{ab}^{\rm CFT}\rangle =ph_{ab}+\rho u_a u_b -2\eta\sigma_{ab}-2\xi \theta^y h_{ab},
\ee
with
\be
p=r_H^{d+1};\qquad \rho =d\, r_H^{d+1};\qquad \eta =r_H^d;\qquad \xi=0.
\ee

\subsection{UV field theory interpretation}

The Dirichlet boundary condition on a finite cutoff hypersurface $\Sigma_c$ necessarily leads to a non-Dirichlet boundary condition at the boundary. The aim of this section is to explore the interpretation of the fluid on the cutoff surface as a state in a deformation of the ultraviolet conformal field theory on the boundary.
The strategy is to look at the solution (\ref{tracegammasol2}-\ref{gammasol2}) near the spacetime boundary  by sending $r\rightarrow\infty$, changing to Fefferman-Graham coordinates. Using the standard AdS/CFT dictionary one can thereby show how the original UV CFT has been deformed. 

The black brane metric in Eddington-Finkelstein coordinates up to order one in the hydrodynamic expansion is 
\ba\label{blackbranemetric3}
ds^2&=&-2\frac{u_a}{\sqrt{h_c}}dy^adr+G_{ab}(y,r)dy^a dy^b,\\
G_{ab}^{(0)}(y,r)&=&r^2\left(h_{ab}-h/h_c\,u_au_b\right),\nn\\
G_{ab}^{(1)}(y,r)&=&\alpha^{(1)}\frac{1}{d}\theta u_au_b+2\beta^{(1)}a_{(a}u_{b)}+\nn\\
&& 
+\tilde{\gamma}^{(1)}\sigma_{ab}+ \gamma^{(1)}\frac{1}{d}\theta h_{ab},\nn
\ea
where the coefficients in $G_{ab}^{(1)}$ are given by (\ref{tracegammasol2}-\ref{gammasol2})  with the obvious notational redefinition 
\be
\alpha^{(1)}\rightarrow\alpha^{(1)}\frac{1}{d}\theta;\quad \beta^{(1)}_a\rightarrow \beta^{(1)}a_{a};\quad \tilde{\gamma}^{(1)}_{ab}\rightarrow\tilde{\gamma}^{(1)}\sigma_{ab},
\ee
and we have again set $L=1$ for convenience.
This metric can be rewritten in the Fefferman-Graham form
\be\label{FGmetric}
ds^2=\frac{d\rho^2}{\rho^2}+\frac{1}{\rho^2}g(z,\rho)_{ab}dz^adz^b,
\ee
by requiring the following transformation equations to be satisfied order by order for the variables $\rho(r,y)$, $z^a(r,y)$ and the metric $g_{ab}(\rho,z)$
\ba\label{changevariables}
\left(\partial_r \rho\right)^2+g_{ab}(z,\rho)\partial_r z^a\partial_r z^b&=&0,\\
\left(\partial_r\rho \right)\left(\partial_a \rho\right)+g_{cd}(z,\rho)\partial_r z^c\partial_a z^d&=&-\frac{u_a}{\sqrt{h_c}}\rho^2,\nn \\
\left(\partial_a \rho\right)\left(\partial_b \rho\right)+g_{cd}(z,\rho)\partial_a z^c\partial_b z^d&=&\rho^2 G_{ab}(y,r). \nn
\ea 
At order zero the change of variables to bring the  metric (\ref{blackbranemetric3}) into Fefferman-Graham form (\ref{FGmetric})
with
\ba\label{zerog}
g(z,\rho)_{ab}^{(0)}&=&A(\rho)\left(h_{ab}-\frac{h(r(\rho))}{h_c}u_au_b\right),\nn\\
A(\rho)&=&\left(1+\frac{1}{4}(r_H\,\rho )^{d+1}\right)^{\frac{4}{d+1}},
\ea
is
\be\label{transformation}
r^{(0)}(\rho) =\frac{1}{\rho}\sqrt{A(\rho)};\qquad
y^{(0)a}(z,\rho)=z^a-\,k(r(\rho))u^a,
\ee
with 
\ba
k(r)&=&\frac{\sqrt{h_c}}{r}\, _2F_1\left(1,\frac{1}{d+1},1+\frac{1}{d+1},\frac{r_H^{d+1}}{r^{d+1}}\right),\nn\\
\partial_r k(r)&=&-\frac{\sqrt{h_c}}{r^2 h(r)}.
\ea
The inverse transformation is given by
\ba\label{inversetransformation}
&&\rho =\frac{1}{r}\left(\frac{2}{1+\sqrt{h(r)}}\right)^{\frac{2}{d+1}}=\frac{1}{r}\sqrt{A(\rho(r))},\nn\\
&&\qquad z^a(y,r)=y^a+\,k(r)u^a
\ea
At first order we can solve equations (\ref{changevariables}) using  the following ansatz
\ba\label{firstansatz}
\rho(y,r)&=&\rho^{(0)}(y,r)+\rho^{(1)}(y,r), \\
z^a(y,r)&=&z^{(0)a}(y,r)+z^{(1)a}(y,r)=\nn\\
&=&z^{(0)a}(y,r)+\frac{1}{d}\theta (y) \,l(r)u^a(y)+m(r)a^a(y),\nn\\
g_{ab}( z,\rho)&=&g_{ab}^{(0)}( z,\rho)+g_{ab}^{(1)}( z,\rho).\nn
\ea
The zeroth order metric (\ref{zerog}) is both dependent on the Fefferman-Graham coordinates $(z,\rho )$ and implicitly dependent on the Eddington-Finkelstein coordinate $y$ through the dependence on $r_H(y)$:
\be
g(z,\rho)_{ab}^{(0)}=A(y,\rho)\left(h_{ab}(z)-\frac{h(y,r(\rho))}{h_c(y)}u_a(z)u_b(z)\right).
\ee
The velocities and the metric can be Taylor expanded as follows
\ba
&&u^{a}(z)=u^{a}(y)+ka^a(y),\\
&&g_{ab}(z,\rho)^{(0)}=g_{ab}^{(0)}(y,r)+u^c(y)k(r)\partial_{c}g_{ab}(y,r)+\nn\\
&&\qquad \qquad \qquad +(\partial_r\rho^{(0)})^{-1}\rho^{(1)}(y,r)\partial_{r}g_{ab}^{(0)}(y,r)=\nn\\
&&\qquad \qquad\quad  =A(\rho(r))\left(h_{ab}-\frac{h}{h_c}u_au_b\right)+\nn\\
&&\qquad \qquad+2\,k(r)A(\rho(r))\left(1-\frac{h}{h_c}\right)a_{(a}u_{b)}+\dots,\nn
\ea
where the last term in the metric
can be ignored since we are ultimately interested in $g_{ab}(y,\rho)$\footnote{The hydrodynamic expansion has been performed in transverse coordinates $y$ and not in the Schwarzschild-type coordinates $z$.}, and we will have to perform the inverse transformation.

We use (\ref{derhor}) and the following derivatives
\ba
\partial_r z^a(y,r)&=&k'(r,y)u^a+l'(r)\frac{1}{d}\theta \,u^a+m'(r)a^a,\nn\\
\partial_c z^a(y,r)&=&\delta_c^a +\partial_c k(y,r)\, u^a+k(y,r)\partial_c u^a,\nn\\
\partial_c k(y,r)&=&\Big( -k(y,r) +\frac{\sqrt{h_c}}{r\,h}+\nn\\
&& -\frac{1}{2}(d+1)\frac{(1-h_c)}{h_c}k(y,r)\Big)\frac{\partial_c r_H}{r_H},\nn\\
\partial_r \rho^{(0)}(y,r)&=&-\frac{1}{r^2\sqrt{h}}\sqrt{A(y,\rho(r))},\nn\\
\partial_c\rho^{(0)}(y,r)&=&\frac{1}{r}\sqrt{A(y,\rho(r))}\frac{(1-\sqrt{h})}{\sqrt{h}}\frac{\partial_c r_H}{r_H},
\ea
where primes denote as usual derivatives with respect to the radial coordinate.
After some manipulations the transformation to Fefferman-Graham form (\ref{firstansatz}) is given by
\ba
l'(r)&=&\frac{1}{r}\sqrt{\frac{h_c}{h}}k'+k'k-2k'\frac{1}{r}\frac{\sqrt{h_c}}{h}+\\
&&+\frac{1}{2}k'k(d+1)\frac{(1-h_c)}{h_c}+\frac{1}{r^2}k'\frac{h_c}{h}\alpha^{(1)},\nn\\
m'(r)&=&\frac{1}{r}k'\frac{(1-\sqrt{h_c})}{\sqrt{h_c}}\delta (r_c)+\frac{1}{r^2}k'\beta^{(1)}+kk',\nn
\ea
and after redefining
\be
\rho^{(1)}(y,r)=\frac{1}{r}\sqrt{A}\, X^{(1)}(y,r)\frac{1}{d}\theta
\ee
we have
\be
X^{(1)\prime}(y,r)=k'\left(\frac{1}{2r}\sqrt{\frac{h_c}{h}}\alpha^{(1)}-\frac{(1-\sqrt{h})}{\sqrt{h}}\right),
\ee
which gives
\be\label{Xsol}
X^{(1)}(y,r)=k+\frac{1}{r\sqrt{h_ch}}\left(\frac{r_c^d\, r}{r_H^{d+1}}(1-h_c)-1\right).
\ee
The black brane metric up to order one in Fefferman-Graham coordinates can be parametrized by
\ba
g_{ab}^{(1)}(y,\rho)&=&\phi(y,r(\rho))\frac{1}{d}\theta u_au_b+\psi(y,r(\rho))\frac{1}{d}\theta h_{ab}+\nn\\
&&+ \Gamma(y,r(\rho))\sigma_{ab}+2\,\Sigma(y,r(\rho))a_{(a}u_{b)},
\ea
with
\ba\label{solFG}
\phi(y,r)&=&2\,A\,k\frac{h}{h_c} -2\frac{A}{r}\frac{1}{\sqrt{h_c}}+A\,k\, h \frac{(1-h_c)}{h_c^2}(d+1)+\nn\\
&&+\frac{1}{r^2}A\,\alpha^{(1)}-2\,A\,X^{(1)}\frac{h}{h_c},\nn\\
\psi(y,r)&=&-2\,A\,k+\frac{1}{r^2}A\,\gamma^{(1)}+2\,A\,X^{(1)},\nn\\
\Sigma(y,r)&=&+A\, k\frac{h}{h_c}(1-\delta(r_c))+\frac{1}{r}\frac{A}{\sqrt{h_c}}\delta(r_c)+\nn\\
&&-\frac{1}{2}A\,k\,h\frac{(1-h_c)}{h_c^2}(d+1)\delta(r_c)+\frac{1}{r^2}A\,\beta^{(1)},\nn\\
\Gamma(y,r)&=&-2\,A\,k+\frac{A}{r^2}\tilde{\gamma}^{(1)}. 
\ea
Notice that all the above expressions reduce to the expressions given earlier in \cite{Kanitscheider:2009as} using the shifts  (\ref{redefinition}) and then sending $r_c\rightarrow\infty$.

Using the zeroth order expression (\ref{transformation}) into 
(\ref{solFG}), using (\ref{tracegammasol2}-\ref{gammasol2}) and  expanding the metric near the boundary of the spacetime, we can read off the leading radial term in the boundary metric 
\ba \label{def}
g_{ab}^{(0)}(y)&=&\eta_{ab}+(1-1/h_c)u_au_b +\\
&&-\frac{2}{r_c\, h_c^{3/2}}\frac{1}{d}\theta \,u_au_b+\frac{\delta(r_c)}{r_c h_c^{3/2}} a_{(a}u_{b)}+\nn\\
&&-2\left(k(r_c)+\frac{1}{(d+1)}\frac{\sqrt{h_c}}{r_H}\ln h_c\right)\sigma_{ab} + \cdots, \nn
\ea
where the ellipses denote terms with higher powers of derivatives.
Notice that after sending $r_c\rightarrow \infty$ the flat metric on the boundary is restored.

The precise interpretation of the fluid on the cutoff hypersurface in terms of the dual CFT can now be given using \eqref{def}. The metric $g^{(0)}_{ab}$ characterises the background metric for the dual CFT or, equivalently, the source for the CFT stress energy tensor. Whenever $r_c$ is finite and hence $h_{c} \neq 1$, the metric $g^{(0)}_{ab}$ is not flat. 
Imposing a Dirichlet boundary condition on the finite cutoff surface $\Sigma_c$ therefore translates into making a specific deformation of the original UV CFT: a non-flat background metric for the field theory fluid. Hence in terms of the UV CFT the fluid lives in a dynamical background metric, namely a metric which depends on the fluid velocity and temperature; a similar interpretation was given in \cite{Brattan:2011my}.

Note that this deformation in the background metric appears to arise already at zeroth order in the hydrodynamic expansion. This was to be expected because, as discussed around \eqref{form1} and \eqref{energy}, we started from a seed metric in which the time Killing vector is normalised to one at the cutoff hypersurface which implies that its norm at infinity is not canonical. 
One can therefore rescale the coordinates so that the zeroth order term in \eqref{def} is flat; this is achieved by rescaling the direction parallel to the velocity by a factor of $\sqrt{h_c}$ but leaving the directions perpendicular to the velocity unchanged. In terms of these rescaled coordinates, the background metric for the dual theory then becomes
\ba \label{def2}
g_{ab}^{(0)} &=& \eta_{ab} -\frac{2}{r_c\, h_c^{1/2}}\frac{1}{d}\theta \,u_au_b+\frac{\delta(r_c)}{r_c h_c} a_{(a}u_{b)}+\\
&&-2\left(k(r_c)+\frac{1}{(d+1)}\frac{\sqrt{h_c}}{r_H}\ln h_c\right)\sigma_{ab} + \cdots\nn,
\ea
where the ellipses denote terms with higher powers of derivatives.

\section{Connection to the membrane paradigm} \label{six}

Consider the limit in which the cutoff hypersurface $\Sigma_c$ approaches the horizon $r_c\rightarrow r_H$.
In the example of a negative cosmological constant considered above the dual fluid has in this limit, as one would expect, the same properties of the Rindler fluid found in \cite{Compere:2012mt}.
In particular we can show this by taking the limit $r_c \rightarrow r_H$ limit of the fluid stress energy tensor, giving
\be
\eta (r_H)=1,
\ee
with the usual shear over entropy universal bound being satisfied as $s = 1/4\pi$, and the bulk viscosity term vanishing. 

In earlier literature another fluid has been associated to a hypersurface close to the horizon, namely the stretched horizon, within the so-called membrane paradigm, \cite{Damour:1979,Damour:1982,Price:1986}.
The latter is an interpretation of the dynamics of a horizon in terms of a dissipative membrane fluid endowed with specific transport properties.
However the two approaches are quite different even though both  fluids are associated with a hypersurface close to a horizon.
On the one hand, within the fluid/gravity duality approach, we have constructed a hydrodynamic solution using a derivative expansion. For consistency,  Gauss-Codazzi equations representing the conservation of Brown-York stress tensor have to be satisfied. We have seen in (\ref{prescription}) that (up to appropriate conformal factors) the Brown-York stress tensor can be given a holographic interpretation as the stress energy tensor of the dual fluid associated with the finite cutoff hypersurface. The stress energy tensor has to be conserved at arbitrarily high order in the derivative expansion and therefore encodes all the transport coefficients. 

On the other hand the membrane paradigm approach 
does not employ any derivative expansion around an equilibrium solution. Instead the membrane fluid equations are derived by reshuffling the Einstein equations and giving an ad hoc interpretation of them in terms of non-relativistic dissipative  Navier-Stokes equations.

In order to understand better the differences between the two approaches we will explore the membrane paradigm point of view in this section and derive the Damour-Navier-Stokes equation together with additional evolution equations for geometrical quantities defined on a $d$-dimensional spacelike foliation of a $(d+1)$-dimensional hypersurface embedded in a $(d+2)$-dimensional bulk spacetime.
We will follow closely the approach of
\cite{Gourgoulhon:2005ch} by considering the case
 of a general hypersurface foliating the bulk spacetime, with the hypersuface not necessarily null and not necessarily close to the horizon; we then take the near horizon limit at the end.

\subsection{Spacelike foliation of a generic hypersurface $\HH$}

Consider  a generic $(d+1)$-dimensional spacelike, timelike or null\footnote{In this formalism the hypersurface is general, however for our purposes we will later restrict it to be timelike.}  hypersurface $\HH$  and further foliate it by $d$-dimensional closed, i.e.  compact without boundary, spacelike hypersurfaces $S_{\tau}$ so that $\HH=\bigcup_{\tau\in \RR} S_{\tau}$.
This means that the  metric $q$ induced by the spacetime metric $g$ onto $S_{\tau}$ is positive definite and in components can be expressed 
\be\label{projector}
q_{\alpha\beta}=g_{\alpha\beta}
+l_{\alpha}k_{\beta}+k_{\alpha}l_{\beta},
\ee
where $(l,k)$ is a pair of null vectors normal to  $S_{\tau}$ satisfying
\be
l\cdot l=0;\quad k\cdot k=0;\quad l\cdot k=-e^{\sigma},
\ee
where $\sigma$ is a positive number. It is possible to define a unique pair $(l,k)$ by requiring
\be\label{hvector}
h=l-Ck;\quad l\cdot k=-1,
\ee 
where $h$ is the evolution vector tangent to $\HH$ and orthogonal to $S_{\tau}$ at any point in $\HH$ with the property
\be\label{Lieh}
\LL_h {\tau}=h^{\mu}\partial_{\mu}\tau =1;\quad h\cdot h=2C.
\ee
$\LL_h$ can be viewed as the  evolution operator along $\Sigma_c$:
given an infinitesimal displacement described by the parameter $\delta\tau$, each point of $S_{\tau}$ is displaced into $S_{\tau +\delta\tau}$ by the vector $\delta\tau h$, and 
$S_{\tau}$ are then hypersurfaces Lie dragged by the  evolution vector $h$.
The character  of $h$ gives the character of the hypersurface $\HH$, in particular if $C<0$, then $\HH$ and $h$ are timelike.

The vectors $(l,k)$ are co-linear to the pair $(\tilde{l},\tilde{k})$ associated to the dual-null foliation formalism \cite{Hayward:1993wb}
\be\label{tilderel}
l=A\tilde{l};\quad k=B\tilde{k},
\ee
where $(\tilde{l},\tilde{k})$ can be defined as the null normal vectors to two families\footnote{These two families can be uniquely defined if the hypersurface $\HH$ is spacelike or timelike, but there is an arbitrariness in the case $\HH$ is null. } of null hypersurfaces generated by outgoing and ingoing light rays  orthogonally from $S_{\tau}$
\be\label{lu}
\underline{\tilde{l}}=-du;\quad\underline{\tilde{k}}=-dv,
\ee
with $u$ and $v$ being  parameters defining the dual-null foliations.

A natural additional vector that can be constructed out of $(l,k)$ is
\be\label{mvector}
m=l+Ck;\quad m\cdot m=-2C; \quad m\cdot h=0.
\ee
Notice that if $h$ is timelike, $m$ is necessarily spacelike and it defines the normal vector to the hypersurface $\HH$, which, together with $h$ spans the orthogonal space to $S_{\tau}$.

Given a generic tensor $T$ on the bulk spacetime, one can canonically define another tensor of the same covariance type $q^*T$ using the projector
\be
(q^*T)^{\alpha _1\dots\alpha_m}_{\beta _1\dots\beta _n}:=q^{\alpha _1}_{\mu _1}\dots q^{\alpha _m}_{\mu _m}q^{\nu _1}_{\beta _1}\dots q^{\nu _n}_{\beta _n}T^{\mu _1\dots\mu _m}_{\nu _1\dots\nu _n}.
\ee
A tensorial field $T$ for which $q^{*}T=T$ is said to be tangent to the surface $S_{\tau}$.
For any such tangent tensorial fields $T$
we can define the covariant derivative on the spacelike surfaces $S_{\tau}$
\be \label{derivative}
\DD T=q^*\nabla T,
\ee
where clearly $\nabla$ is the covariant derivative on the $(d+2)$-dimensional bulk spacetime.

\subsection{Extrinsic geometry  of the  spacelike surface $S_{\tau}$:}
Given a vector field $v$ orthogonal to the spacelike surface $S_{\tau}$, the deformation tensor along this field is defined as
\begin{equation}\label{shape}
\Theta ^{(v)}:=q^{*}(\nabla v);\qquad  \Theta^{(v)}_{\alpha\beta}=q_{\alpha}^{\nu} q_{\beta}^{\mu}\nabla_{\nu} v_{\mu},
\end{equation}
measuring the variation of the metric in $S_{\tau}$ when the surface $S_{\tau}$ is displaced along $v$. The deformation tensor is symmetric and can be decomposed into a traceless symmetric shear tensor and a trace part
\be\label{shear}
\sigma_{\alpha\beta}^{(v)}:=\Theta^{(v)}_{\alpha\beta}-\frac{1}{d}\theta^{(v)}q_{\alpha\beta};\quad \theta^{(v)}:=q^{\mu\nu}\Theta^{(v)}_{\mu\nu}, 
\ee
the latter measuring the change of the area element in $S_{\tau}$ when displaced by $v$.


The variation of the normal fields to $S_{\tau}$ with respect to each other is instead contained in the normal fundamental forms, which for the pair $(l,k)$ can be written as
\ba
\Omega^{(l)}:=\frac{1}{k\cdot l}k\cdot\nabla_{q}l \,;&\quad& \Omega_{\alpha}^{(l)}=\frac{1}{k_{\rho} l^{\rho}}k_{\mu}q^{\nu}_{\alpha}\nabla\nud l\muu;\label{normal1} \\
\Omega^{(k)}:=\frac{1}{k\cdot l}l\cdot\nabla_{q}k \,;&\quad& \Omega_{\alpha}^{(k)}=\frac{1}{k_{\rho} l^{\rho}}l_{\mu}q^{\nu}_{\alpha}\nabla\nud k\muu,\label{normal2} 
\ea
with the important relation\footnote{For completeness we have reported the general equation, however since we will always use the definition (\ref{hvector}), we can set $\sigma=0$.}
\be\label{relation}
\Omega^{(k)}=-\Omega^{(l)}+\DD \sigma.
\ee
The extrinsic curvature can be defined out of the shape tensors along $k$ and $l$ as follows
\be
\KK_{\beta\gamma}^{\alpha}=k^{\alpha}\Theta_{\beta\gamma}^{(l)}+l^{\alpha}\Theta_{\beta\gamma}^{(k)}.
\ee
The extrinsic curvature  together with the normal fundamental forms are the sufficient
quantities to describe the extrinsic geometry of the spacelike surfaces $S_{\tau}$.

\subsection{Kinematics of the spacelike surfaces $S_{\tau}$:}
Covariant derivatives of the normal vectors $l$ and $k$ can be decomposed using  quantities
defined on the spacelike surfaces $S_{\tau}$ after using the projector (\ref{projector})  
\ba
\nabla_{\alpha} l_{\beta}&=&\Theta^{(l)}_{\alpha\beta}+\Omega^{(l)}_{\alpha}l_{\beta}-l_{\alpha}\nabla_k l_{\beta}-\nu_{(l)}k_{\alpha}l_{\beta},\label{cov1}\\
\nabla_{\alpha} k_{\beta}&=&\Theta^{(k)}_{\alpha\beta}-\Omega^{(l)}_{\alpha}k_{\beta}-k_{\alpha}\nabla_l k_{\beta}-\nu_{(k)}l_{\alpha}k_{\beta},\label{cov2}
\ea
where we have used equations (\ref{tilderel}-\ref{lu}) to derive 
\ba
\nabla_l l=\nu_{(l)}l;&\quad&\nu_{(l)}:=\LL_{l}\ln A;\\
\nabla_k k=\nu_{(k)}k;&\quad&\nu_{(k)}:=\LL_{k}\ln B.
\ea
Combining (\ref{cov1}-\ref{cov2}) and using the definitions (\ref{hvector}) and (\ref{mvector}) we can derive the covariant derivatives for $m$ and $h$
\ba
\nabla_{\alpha}m_{\beta}&=&\Theta_{\alpha\beta}^{(m)}+\Omega_{\alpha}^{(l)}h_{\beta}-l_{\alpha}\nabla_kl_{\beta}-\nu_{(l)}k_{\alpha}l_{\beta}+\nn\\
&&+(\nabla_{\alpha} C)k_{\beta}-Ck_{\alpha}\nabla_lk_{\beta}-C\nu_{(k)}l_{\alpha}k_{\beta},\label{covm}\\
\nabla_{\alpha}h_{\beta}&=&\Theta_{\alpha\beta}^{(h)}+\Omega_{\alpha}^{(l)}m_{\beta}-l_{\alpha}\nabla_kl_{\beta}-\nu_{(l)}k_{\alpha}l_{\beta}+\nn\\
&&-(\nabla_{\alpha} C)k_{\beta}+Ck_{\alpha}\nabla_lk_{\beta}+C\nu_{(k)}l_{\alpha}k_{\beta}\label{covh},
\ea
where we used
\be
\Theta^{(m)}=\Theta^{(l)}+C\Theta^{(k)};\quad \Theta^{(h)}=\Theta^{(l)}-C\Theta^{(k)},
\ee
which follow straightforwardly from definition (\ref{shape}).

\subsection{Dynamics of the spacelike surfaces $S_{\tau}$}

Having derived the expressions for the first derivatives of the vector fields $m$ and $h$, we can now manipulate the second order derivatives using
the Ricci identity, which for a generic  vector field $v$ is written as 
\be\label{ricciidentity}
R^{\gamma}_{\,\,\rho\mu\nu}v^{\rho}=(\nabla_{\mu}\nabla_{\nu}-\nabla_{\nu}\nabla_{\mu})v^{\gamma},
\ee
to derive evolution equations along $m$ and $h$ for the quantities defined on the spacelike hypersurfaces $S_{\tau}$. Moreover, requiring the metric to be a solution of Einstein equations one can always trade the Ricci tensor for the trace subtracted bulk stress tensor 
\be\label{einsteineq}
R_{\mu\nu}=\bar{\TT}_{\mu\nu},
\ee
where we have used the same notation as in the earlier sections. 
There are as many equations as there are many independent ways to project the Einstein equations along $m$, $h$ and $q$. Some of the equations are listed below for illustration and we refer to
 appendix  \ref{A2} for details of their derivation. 
Projection along $q$ and $m$ results in the generalized Damour-Navier-Stokes equation as obtained previously in \cite{Gourgoulhon:2005ch}
\ba\label{eqq1}
&&q^*\LL_h\Omega_{\alpha}^{(l)}+\theta^{(l)}\Omega^{(l)}_{\alpha}=-\DD_{\mu}\sigma_{\alpha}^{(m)\mu}+\frac{(d-1)}{d}\DD_{\alpha}\theta^{(m)}+\nn\\
&&\qquad  +\DD_{\alpha}\left(\nu_{(l)}+C\nu_{(k)}\right)-\theta^{(k)}\DD_{\alpha}C+q^*\bar{\TT}_{\mu\alpha}m^{\mu},
\ea
which describes the evolution of the normal fundamental tensor $\Omega^{(l)}_{\alpha}$ along the vector field $h$. Other equations can be considered, e.g. the
evolution equation for $\theta^{(m)}$ along $m$ be obtained projecting twice along $m$
\ba\label{eqq2}
&&\nabla_{m}\theta^{(m)}+\nabla_m\left(\nu_{(l)}+C\nu_{(k)}\right)+\Theta_{\alpha\beta}^{(m)}\Theta^{(m)\alpha\beta}+\nn\\
&&-\nabla_l\nu_{(l)}-\nu_{(l)}\theta^{(l)}-2C\Omega^{(l)\alpha}\DD_{\alpha}\ln A-(\theta^{(k)}+\nu_{(k)})\nabla_mC+\nn\\
&&-C\nabla_{\alpha}\left(\nabla_kl^{\alpha}+\nabla_lk^{\alpha}\right)+4C\Omega^{(l)\alpha}\Omega^{(l)}_{\alpha}-2C\nu_{(l)}\nu_{(k)}+\nn\\
&&-C^2\nu_{(k)}\theta^{(k)}-C^2\nabla_k\nu_{(k)}+\bar{\TT}_{\mu\nu}m^{\mu}m^{\nu}=0, 
\ea
or the evolution equation of $\theta^{(m)}$ along $h$ by projecting along $h$ and $m$
\ba\label{eqq3}
&&\LL_{h}\theta^{(m)}+\LL_h\left(\nu_{(l)}+C\nu_{(k)}\right)+\Theta_{\alpha\beta}^{(h)}\Theta^{(m)\alpha\beta}+\nn\\
&&-\nabla_l\nu_{(l)}-\nu_{(l)}\theta^{(l)}-(\theta^{(k)}+\nu_{(k)})\LL_hC+ \\
&&+C\nabla_{\alpha}\left(\nabla_kl^{\alpha}-\nabla_lk^{\alpha}\right)+C\nu_{(k)}\theta^{(k)}+\nn\\
&&+C^2\nabla_k\nu_{(k)}+\bar{\TT}_{\mu\nu}m^{\mu}h^{\nu}=0. \nn
\ea
An example of a tensorial equation is the evolution equation of the shape tensor $\Theta_{\alpha\beta}^{(m)}$ which is
\ba\label{eqq4}
q^*\LL_h\Theta_{\alpha\beta}^{(m)}&=&\Theta_{\alpha\mu}^{(m)}
\Theta^{(h)\mu}_{\beta}+\nu_{(l)}\Theta_{\alpha\beta}^{(l)}
+C\Omega_{\alpha}\DD_{\beta}\ln A+\nn\\
&&+\Theta_{\alpha\beta}^{(k)}\LL_hC-C\nabla_{\mu}\left(\nabla_kl_{\nu}-\nabla_lk_{\nu} \right)q^{\mu}_{\alpha}q^{\nu}_{\beta}+\nn\\
&&-C^2\nu_{(k)}\Theta_{\alpha\beta}^{(k)}+h^{\nu}R_{\rho\mu\nu\sigma}m^{\mu}q^{\sigma}_{\alpha}q^{\rho}_{\beta}.
\ea

\subsection{The null limit}  
The equations derived above are valid for a generic hypersurface $\HH$. The limit in which the hypersurface $\HH$ is null can be achieved by sending
\be
 C\rightarrow 0;\quad m\rightarrow l;\quad h\rightarrow l.
\ee
 In this limit the null vector field $l$ plays the double role of the tangent and the normal to the null hypersurface $\HH$. In this way the system of equations derived above becomes degenerate and reproduces the usual equations for the dynamics of null hypersurfaces, see e.g. \cite{Gourgoulhon:2005ng}.
Equation (\ref{eqq1}) becomes the Damour-Navier-Stokes equation
\ba\label{eqq1bis}
q^*\LL_l\Omega_{\alpha}^{(l)}+\theta^{(l)}\Omega^{(l)}_{\alpha}&=&-\DD_{\mu}\sigma_{\alpha}^{(l)\mu}+\frac{(d-1)}{d}\DD_{\alpha}\theta^{(l)}+\nn\\
&&+\DD_{\alpha}\nu_{(l)}+q^*\bar{\TT}_{\mu\alpha}l^{\mu},
\ea
while equations (\ref{eqq2}-\ref{eqq3}) become the so-called null Raychaudhuri equation
\be\label{eqq2bis}
\nabla_{l}\theta^{(l)}+\Theta_{\alpha\beta}^{(l)}\Theta^{(l)\alpha\beta}-\nu_{(l)}\theta^{(l)}+\bar{\TT}_{\mu\nu}l^{\mu}l^{\nu}=0,
\ee
and the tensorial equation (\ref{eqq4}) becomes the tidal force equation for $\Theta_{\alpha\beta}^{(l)}$
\be\label{eqq4bis}
q^*\LL_l\Theta_{\alpha\beta}^{(l)}=\Theta_{\alpha\mu}^{(l)}\Theta^{(l)\mu}_{\beta}+\nu_{(l)}\Theta_{\alpha\beta}^{(l)}+q^*(l^{\nu}R_{\beta\mu\nu\alpha}l^{\mu}).
\ee
The essence of the gravitational membrane paradigm is to  reinterpret equation (\ref{eqq1bis}) formally in terms of a non relativistic dissipative Navier-Stokes equation
\be
\LL_{l}\PP_{\alpha}+\theta \PP_{\alpha}=-\partial_{\alpha}p+2\eta \DD_{\beta}\sigma^{(l)\beta}_{\alpha}+\xi \partial_{\alpha}\theta-f_{\alpha}
\ee
upon the identification 
\ba
\PP_{\alpha}=-\Omega^{(l)}_{\alpha};&\qquad& \nu_{(l)}=p; \quad\eta=1/2;\nn\\
\xi=-(d-1)/d;&\qquad& q^*\bar{\TT}_{\mu\alpha}l^{\mu}=f_{\alpha},
\ea
$\PP_{\alpha}$ being the momentum surface density, $p$ the fluid pressure, $\eta$ and $\xi$ the shear and bulk viscosity associated respectively with the shear tensor $\sigma_{\alpha\beta}^{(l)}$ and the expansion rate $\theta^{(l)}$, and $f_{\alpha}$ an external force surface density. 
Notice that the membrane fluid does not appear to be physical since it has a negative bulk viscosity. 

\subsection{Relation to the holographic fluid}

This putative membrane fluid is not directly related to the near horizon Rindler fluid
discussed in previous sections obtained from a holographic prospective, and here we depart from the interpretation given in \cite{Emparan:2013ila}. Let us summarise the differences between the two setups. In the membrane approach we consider a generic timelike hypersurface for which we can take a null limit and we do not need to input specific information about the horizon nature of the null hypersurface. In the holographic approach we consider a timelike hypersurface outside a horizon; we fix the metric on this hypersurface. 

In the membrane approach we work in a manifestly non-relativistic formulation, in that we consider the $(d+1)$-dimensional hypersurfaces to be foliated by spacelike $d$-dimensional hypersurfaces. The normal vectors to these spacelike hypersurfaces play an important role: it is their variation with respect to each other (a normal fundamental form) which is interpreted as the momentum density of the membrane fluid. Moreover, the shear and expansion of the spacelike surfaces along normal vectors is interpreted as the shear and expansion of the membrane fluid. 

In the fluid/gravity approach, we can work either in relativistic or non-relativistic formulations. In the earlier sections of this paper we have used the relativistic approach as this is computationally more efficient but one could equally work with non-relativistic expansions as in \cite{Bredberg:2011jq, Compere:2011dx}. The holographic fluid properties are captured by the induced metric on the $(d+1)$-dimensional timelike hypersurface, which acts as the background metric for the fluid, and by the extrinsic curvature of the hypersurface, which encodes the pressure, energy density and velocity of the fluid. 

The putative membrane fluid is generally dissipative and the shear and bulk viscosity take universal values. 
Note however that if the chosen spacetime metric is what we have called a zeroth order equilibrium metric
the shear tensor and the expansion rate of the horizon would vanish, so the dissipative terms would disappear.

In the holographic interpretation of fluid/gravity duality the zeroth order gravity solution represents a thermodynamic equilibrated state in which the behaviour is non-dissipative.
In order to get dissipative behaviour at all it is necessary to perturb the thermodynamic solutions around equilibrium. This means that the existence of a family of solutions close to equilibrium thermodynamic solutions is an essential requirement to obtain hydrodynamic behaviour. The shear and bulk viscosity depend on the system under consideration although $\eta/s$ takes the well-known universal value of $1/4\pi$  in Einstein gravity. 

In the fluid/gravity approach the Navier-Stokes equations (together with specific corrections) arise from working out the Einstein equations in a gradient expansion. All components of the Einstein equations are needed to work out the fluid equations. In the membrane paradigm only a specific projection of the Einstein equations is used to obtain the membrane fluid equations: remaining equations such as the tidal force equation do not have a natural interpretation in terms of fluid quantities. 


\section{Conclusions} \label{seven}

In this paper we have presented a construction
of generic $(d+2)$-dimensional near equilibrium metrics corresponding to the  hydrodynamic regime of putative $(d+1)$-dimensional holographic fluids associated with timelike hypersurfaces foliating a general (i.e. general bulk stress energy tensor) bulk spacetime.
Using the method of Hamiltonian holographic renormalisation we gave a prescription for the fluid stress energy tensor in the case of (conformally) flat Dirichlet boundary conditions on these timelike hypersurfaces. Our prescription is consistent with standard holographic results in the limit where the timelike hypersurface is taken to the conformal boundary.

The resulting stress tensor is proportional to the Brown-York stress tensor of the corresponding hypersurface plus certain boundary terms. These boundary terms are in principle uniquely defined when the hypersurfaces are taken to the asymptotic boundary and represent the necessary counterterms to ensure the on-shell action to be finite. On a finite cutoff the above mentioned boundary terms cannot be fixed uniquely but we have shown that in the hydrodynamic regime they only provide a redefinition of the thermodynamic quantities without  affecting the thermodynamic relation nor hydrodynamics.

Another source of boundary terms is  the part of the spacetime between the finite cutoff and the  boundary at infinity. In the spirit of Wilsonian holographic renormalisation \cite{Faulkner:2010jy,Heemskerk:2010hk}, see also \cite{Nickel:2010pr}, this part of the geometry is dual to the contribution of high energy degrees of freedom
which can be integrated out giving rise to a boundary effective action.
In this paper we are taking a hard cutoff point of view along with e.g. \cite{Brattan:2011my,Emparan:2013ila}, namely we consider only the part of the geometry between the horizon and the cutoff itself ignoring everything else beyond it. It would be interesting to see how in the hydrodynamic regime our results can be matched to the local contributions coming from the so called UV part of spacetime.

Relative to earlier works (see for example  \cite{Brattan:2011my,Emparan:2013ila}) 
we have clarified a number subtleties. In particular, we have emphasised the fact that different coordinate systems
give physically distinct fluids on timelike hypersurfaces obtained at any given radial cutoff. At leading order in the hydrodynamic expansions we can simply perform coordinate transformations and relate the pressures and energy densities but the hydrodynamic expansions are taken about different hypersurfaces and in particular with respect to different dual field theory spacetime coordinates; hence out of equilibrium we are dealing with physically different fluids.
In the case of pure AdS gravity this subtlety does not arise with a flat or conformally flat Dirichlet boundary condition on the finite cutoff due to the fact that the conformal factor does not depend on the field theory coordinates but the issues discussed here would be relevant for dealing with hydrodynamics for cases such as AdS R-charged black holes (obtained as decoupling limits of rotating D3-branes)\footnote{See the recent work \cite{Erdmenger:2014jba}.}. 

One of the conclusions of \cite{Brattan:2011my,Emparan:2013ila}) was that the fluid changes from a relativistic to non-relativistic fluid as the radial coordinate decreases. Here we found that the near horizon description is the Rindler fluid of \cite{Compere:2011dx,Compere:2012mt}, which indeed can be viewed as a non-relativistic fluid.

After discussing classes of spacetimes with a general bulk stress tensor at thermodynamic equilibrium, we concentrated on the specific case of Einstein gravity in AdS and verified the consistency of our prescription for the fluid stress tensor with standard holographic results when the timelike hypersurface is taken to the conformal boundary of AdS. Having at our disposal the holographic dictionary at conformal infinity we gave a precise interpretation of the fluid on the cutoff hypersurface in terms of a specific deformation of the UV CFT. The resulting UV fluid can be thought of as living in a non-flat background, depending on the fluid velocity and temperature.  

Finally, we have also explored the near horizon limit of the cutoff AdS fluid, which up to first order in a gradient expansion is effectively a Rindler fluid. We have emphasised the differences with the membrane fluid discussed earlier in the literature departing also from the interpretation of the membrane paradigm given in \cite{Emparan:2013ila}.
In particular we have showed how through the holographic type construction one can obtain fluid equations up to arbitrary order in a gradient expansion in contrast to the membrane fluid which is only a suitable rewriting of certain components of Einstein equations into a fluid-like fashion. 
Of course one could easily use explicitly the form of the horizon metric and some derivative expansion of it in the membrane fluid equations,
but this would be equivalent to taking the holographic point of view and would for example spoil the original structure of the membrane fluid equation.

 Recently an interesting connection between asymptotically flat spacetimes and asymptotically AdS black holes has emerged 
 \cite{Caldarelli:2012hy,Caldarelli:2013aaa}: it has been shown that asymptotically AdS black holes compactified on tori correspond to certain asymptotically flat Schwarzschild black branes and the holographic dictionary for the stress energy tensor has been derived through generalized dimensional reduction. It would be interesting to see how our construction would fit into this framework, and also how 
our construction can be applied to blackfolds \cite{Emparan:2009cs,Emparan:2011hg} which interpolate between asymptotically AdS and asymptotically flat regions.

\section*{Acknowledgments}
NPF is grateful to Paul McFadden, Goffredo Chirco and Michal Heller for discussions.
This work is part of the research program of the Stichting voor
Fundamenteel Onderzoek der Materie (FOM), which is financially
supported by the  Nederlandse Organisatie voor Wetenschappelijk
Onderzoek (NWO). MMT acknowledges support from the KNAW (Royal Netherlands Academy of Arts and Sciences), STFC and via a grant of the John Templeton Foundation. The opinions expressed in this 
publication are those of the authors and do not necessarily reflect the views of the John Templeton Foundation. 

\appendix

\section{Details of the hydrodynamic expansion}\label{appendix1}

The  zeroth order metric is
\be\label{metric10}
ds^2=-\frac{2}{\lambda}u_adx^adr+\left (Gh_{ab}-Fu_au_b\right)dx^adx^b,
\ee
the inverse metric is
\be
g^{(0)rr}=\lambda^2 F;\qquad g^{(0)ra}=\lambda u^a;\qquad g^{(0)ab}=\frac{h^{ab}}{G}.
\ee
The zeroth order Christoffel symbols are 
\ba\label{symbols0}
\Gamma_{rr}^{(0)r}&=&0,\qquad \Gamma_{rr}^{(0)a}=0, \qquad
\Gamma_{ra}^{(0)r}= \frac{1}{2}\lambda F'u_a, \\
\Gamma_{ab}^{(0)r}&=&-\frac{1}{2}\lambda^2 F\left( G'h_{ab}-F'u_au_b\right),\nonumber \\
\Gamma_{rb}^{(0)a}&=&\frac{1}{2}\frac{G'}{G}h^a_b,\qquad 
\Gamma_{ab}^{(0)c}=-\frac{1}{2}\lambda u^c\left(G'h_{ab}-F'u_au_b \right),\nonumber\ea
with useful contractions being
\be
\Gamma_{\mu r}^{(0)\mu}=\frac{1}{G^{d/2}}\partial_rG^{d/2};\qquad \Gamma_{\mu a}^{(0)\mu}=0.
\ee
After promoting the parameters to depend on $x$, the metric (\ref{metric10})
is not solution of the Einstein's equations.
However adding corrections $g^{(n)}$ one can construct a solution  order by order.
The usual gauge choice is $g^{(n)}_{rr}=g^{(n)}_{ra}=0$ so that the lines at constant $x^a$ are bulk radial null geodesics and the metric keeps the Eddington-Finkelstein form to all orders, which is useful in order to avoid coordinate singularities at the horizon.

The variations of the Christoffel symbols can be computed formally  to all orders  using
\be
\delta\Gamma_{\mu\nu}^{(n)\rho}=\frac{1}{2}g^{(0)\rho\lambda}\left( \bar{\nabla}_{\mu}g^{(n)}_{\lambda\nu}+\bar{\nabla}_{\nu}g^{(n)}_{\lambda\mu}-\bar{\nabla}_{\lambda}g^{(n)}_{\mu\nu} \right),
\ee
where $\bar{\nabla}_{\mu}$ is the covariant derivative with respect to the background zeroth order metric.
Hence using (\ref{symbols0}) we can compute  
\ba
\delta\Gamma_{rr}^{(n)\mu}&=&0,\\
\delta\Gamma_{ra}^{(n)r}&=&\frac{1}{2}\lambda u^c\partial_r g_{ca}^{(n)}-\frac{1}{2}\lambda \frac{G'}{G}u^ch^d_ag^{(n)}_{cd}, \nonumber \\
\delta\Gamma_{ab}^{(n)r}&=&-\frac{1}{2}\lambda^2 F\partial_rg_{ab}^{(n)}+\frac{1}{2}\lambda ^2\left( G'h_{ab}-F'u_au_b  \right)u^cu^dg^{(n)}_{cd}, \nonumber \\
\delta\Gamma_{rb}^{(n)a}&=&\frac{1}{2}\frac{1}{G}h^{ac}\partial_r g^{(n)}_{cb}-\frac{1}{2}\frac{G'}{G^2}h^{ac}h^d_bg^{(n)}_{cd}, \nonumber \\
\delta\Gamma_{ab}^{(n)c}&=&-\frac{1}{2}\lambda u^c\partial_rg^{(n)}_{ab}+\frac{1}{2}\frac{\lambda}{G}\left(G'h_{ab}-F'u_au_b\right)h^{ce}u^fg^{(n)}_{ef}, \nonumber \ea
and the useful contractions
\be
\delta\Gamma_{\rho r}^{(n)\rho}=\frac{1}{2}\partial_r\left ( \frac{1}{G}h^{cd}g_{cd}^{(n)}   \right);\qquad \delta\Gamma_{\rho a}^{(n)\rho}=0.
\ee
The variations of the Ricci tensor to all orders can be computed using
\be
\delta R_{\mu\nu}^{(n)}=-\bar{\nabla}_{\mu}\delta \Gamma_{\rho \nu}^{(n)\rho}+\bar{\nabla}_{\rho}\delta \Gamma_{\mu \nu}^{(n)\rho},
\ee
which gives (\ref{Riccivar}).

The Christoffel symbols up to first order obtained from the seed metric (\ref{metric10}) with parameters depending on $x$ are
\ba
\Gamma_{rr}^{(1)\mu}&=&0,\\
\Gamma_{ra}^{(1)r}&=&\frac{1}{2}\lambda u_a F'+\frac{1}{2}a_a-\frac{1}{2}D_a^{\perp}\ln \lambda,\nonumber \\
\Gamma_{ab}^{(1)r}&=&-\frac{1}{2}\lambda^2 F\left(G' h_{ab}-F'u_au_b \right)-\lambda K_{ab}G+u_{(a}a_{b)}\lambda F+\nn\\
&&+u_{(a}\lambda D^{\perp}_{b)}F-\frac{1}{2}\lambda u_{a}u_{b}DF+\nn\\
&&-\frac{1}{2}\lambda h_{ab}DG+\lambda Fu_{(a}\partial_{b)}\ln \lambda,\nonumber \\
\Gamma_{rb}^{(1)a}&=&\frac{G'}{2G}h^a_b+\frac{1}{G\lambda}\Omega^a_b+\frac{1}{2G\lambda}a^au_b-\frac{1}{2\lambda G}u_bD^{\perp a}\ln \lambda,\nonumber \\
\Gamma_{ab}^{(1)c}&=&-u^cK_{ab}+u^cu_{(a}a_{b)}-\frac{1}{2}\lambda u^c\left( G'h_{ab}-F'u_au_b\right)+\nn\\
&&-\frac{(G-F)}{G}\left(2u_{(a}\Omega^c_{b)}+a^cu_au_b\right)+\frac{1}{G}h^c_{(a}\partial_{b)}G+\nn\\
&&-\frac{1}{2G}h_{ab}D^{\perp c}G+\frac{1}{2G}u_au_bD^{\perp c}F+u^cu_{(a}\partial_{b)}\ln \lambda.\nonumber 
\ea
and useful contractions
\ba
\Gamma_{\rho r}^{(1)\rho}&=&\frac{1}{G^{d/2}}\partial_rG^{d/2};\nn\\
\Gamma_{\rho a}^{(1)\rho}&=&\frac{1}{G^{d/2}}\partial_a G^{d/2}+u_aD\ln \lambda -D^{\perp}_a\ln \lambda. 
\ea
The Ricci tensor components are shown in (\ref{Riccifirst}).

The general solution to equations  (\ref{gammatracefirst}-\ref{gammafirst}) is
\ba
\gamma^{(1)}&=&(\gamma_0\, r+\gamma_1\, r^2)\theta,\nn\\
\beta^{(1)}_a&=&r\left(\beta_0 \,r+\frac{\beta_1}{r^d}-\frac{L}{r_c \sqrt{h_c}}\delta(r_c) \right)a_a, \\
\alpha^{(1)}&=&\frac{\alpha_0}{r^{d-1}}\theta +\frac{L}{d \, h_c^{3/2}}\frac{r}{r_c}\left( (d-1)(h-1)+2(h_c-1)\right)\theta,\nn\\
\tilde{\gamma}_{ab}^{(1)}&=&\frac{1}{2}\frac{L\sqrt{h_c}}{r_cr_H}r^2\left(4 k(r)+\tilde{\gamma}_0+\tilde{\gamma}_1\,\ln h(r)\right)\sigma_{ab}, \nn 
\ea
where $k(r)$ is given by (\ref{hyper}) and $\gamma_0$, $\gamma_1$, $\beta_0$, $\beta_1$, $\alpha_0$, $\tilde{\gamma}_0$ and $\tilde{\gamma}_1$ are integration constants.
Dirichlet boundary conditions fix some of them to
\ba 
\gamma_1 &=&-\frac{\gamma_0}{r_c},\nn\\
\beta_1 &=&r_c^d\left( -r_c\beta_0 +\frac{L}{r_c\sqrt{h_c}}\delta(r_c)\right),\\
\alpha_0 &=&-\frac{(d+1)}{d}\frac{Lr_c^{d-1}}{h_c^{3/2}}\left( h_c-1\right),\nn\\
\tilde{\gamma}_0&=&-4k(r_c)-\tilde{\gamma}_1\ln h_c, \nn
\ea
Landau gauge conditions (\ref{landaufirst1}-\ref{landaufirst2}) fix the other integration constants to
\ba
\gamma_0 &=&\frac{2L}{r_c\sqrt{h_c}},\\
\beta_0 &=&-\frac{L\big( d\, h_c(2+(d-1)\delta(r_c))-(d+1)(2+d\,\delta (r_c))\big)}{2(d+1)r_c^2 h_c^{3/2}}\delta(r_c),\nn
\ea 
and requiring regularity on the horizon $r_H$ gives
\be
\tilde{\gamma}_1=\frac{4}{d+1},
\ee
leading to the solutions (\ref{tracegammasol}-\ref{tracegamma}).

\section{Details on the 2+1+1 foliation of the spacetime}\label{A2}
Knowing  (\ref{Lieh}), (\ref{tilderel}) and (\ref{lu}) it is possible to show the following identities 
\be\label{usefulrel2}
\DD_{\alpha}B=0;\qquad \DD_{\alpha}(C/A)=0\quad {\rm on}\quad \HH.
\ee
These can then be used to derive the useful relations 
\be\label{usefulrel1}
q^{\nu}_{\alpha}\nabla_kl_{\nu}=-\Omega_{\alpha}^{(l)}+\DD_{\alpha}\ln A;\qquad q^{\nu}_{\alpha}\nabla_{l}k_{\nu}=\Omega_{\alpha}^{(l)}.
\ee
Let us first consider 
\ba
&&(\nabla_{\beta}\nabla_{\alpha}-\nabla_{\alpha}\nabla_{\beta})m^{\beta}=\nn\\
&&
\quad \nabla_{\beta}\Theta_{\alpha}^{(m)\beta}+h^{\beta}\nabla_{\beta}\Omega_{\alpha}^{(l)}+\Omega_{\alpha}^{(l)}\theta^{(h)}+\Omega_{\alpha}^{(l)}\nu_{(l)}+\nn\\
&&\quad-C\nu_{(k)}\Omega_{\alpha}^{(l)} +\Theta_{\alpha}^{(h)\beta}\Omega_{\beta}^{(l)}-\Theta_{\alpha}^{(l)\beta}\DD_{\beta}\ln A+\nn\\
&&\quad +m_{\alpha}\Omega^{(l)\beta}\Omega_{\beta}^{(l)}-l_{\alpha}\Omega^{(l)\beta}\DD_{\beta}\ln A-l_{\alpha}\nabla_{\beta}(\nabla_kl^{\beta})+\nn\\
&&\quad +m_{\alpha}\nu_{(l)}\nu_{(k)}-k_{\alpha}\LL_{l}\nu_{(l)}-\nu_{(l)}\nabla_{l}k_{\alpha}-\nu_{(l)}^2k_{\alpha}+\nn\\
&&\quad -\nu_{(l)}k_{\alpha}\theta^{(l)}+(\theta^{(k)}+\nu_{(k)})\nabla_{\alpha}C-Ck_{\alpha}\nabla_{\beta}(\nabla_{l}k^{\beta})+\nn\\
&&\quad -\nu_{(k)}l_{\alpha}\LL_kC-Cl_{\alpha}\LL_k\nu_{(k)}-C\nu_{(k)}\nabla_kl_{\alpha}-C\nu_{(k)}^2l_{\alpha}+\nn\\
&&\quad -C\nu_{(k)}\theta^{(k)}l_{\alpha}-\nabla_{\alpha}\theta^{(m)}-\nabla_{\alpha}(\nu_{(l)}+C\nu_{(k)})+\nn\\
&&\quad -\Theta_{\alpha}^{(k)\beta}\DD_{\beta}C+\nu_{(k)}l_{\alpha}\nabla_kC,
\ea
where we repeatedly used (\ref{covm}).
By further projecting on $q$, using 
(\ref{usefulrel1}-\ref{usefulrel2}) and the relations
\ba
q^{\nu}_{\alpha}\nabla_{\mu}\Theta_{\nu}^{(m)\mu}&=&\DD_{\mu}\Theta_{\alpha}^{(m)\mu}+\Theta_{\alpha}^{(m)\mu}\DD_{\mu}\ln A,\\
q^{\nu}_{\alpha}\LL_h\Omega_{\nu}^{(l)}&=&q^{\nu}_{\alpha}h^{\mu}\nabla_{\mu}\Omega_{\nu}^{(l)}+\Theta_{\alpha}^{(h)\mu}\Omega_{\mu}^{(l)},
\ea
as well as Einstein equations  (\ref{einsteineq}) we get
the generalized Damour-Navier-Stokes equation (\ref{eqq1}).
Equations (\ref{eqq2}-\ref{eqq3}) can be derived similarly.

To derive equation (\ref{eqq4}) we consider 
\be
h^{\rho}\nabla_{\rho}\nabla_{\mu}m_{\nu}q^{\mu}_{\alpha}q^{\nu}_{\beta}=h^{\rho}\left( R_{\nu\sigma\rho\mu}m^{\sigma}+\nabla_{\mu}\nabla_{\rho}m_{\nu}\right)q^{\mu}_{\alpha}q^{\nu}_{\beta},
\ee
and repeatedly make use of (\ref{covm}-\ref{covh}), (\ref{usefulrel1}) and
\be
q^{\mu}_{\alpha}q^{\nu}_{\beta}\LL_h\Theta_{\mu\nu}^{(m)}=q^{\mu}_{\alpha}q^{\nu}_{\beta}\nabla_h\Theta_{\mu\nu}^{(m)}+\Theta_{\alpha\rho}^{(m)}\Theta_{\beta}^{(h)\rho}+\Theta_{\beta\rho}^{(m)}\Theta_{\alpha}^{(h)\rho}.
\ee

\newpage

\providecommand{\href}[2]{#2}\begingroup\raggedright
\endgroup


\end{document}